\documentclass[prb,twocolumn,showpacs,preprintnumbers,amsmath,amssymb,floatfix,nofootinbib]{revtex4-2}
\usepackage{graphicx}
\usepackage{dcolumn}
\usepackage{bm}
\usepackage{color}
\usepackage{multirow}
\usepackage{tabularx}
\usepackage{ulem}
\usepackage[version=4]{mhchem} 
\usepackage{hyperref}

\def \E_A{\E_\text{A}}
\def \E_B{\E_\text{B}}

\definecolor{gold}{rgb}{0.85,.66,0}
\definecolor{dark-green}{rgb}{0.17, 0.63, 0.22}
\definecolor{RVP}{RGB}{220, 10, 10}

\begin{document}

\title{Antiferromagnetic nonreciprocity of light emission in CuB$_2$O$_4$} 

\author{A.~R.~Nurmukhametov,$^{1}$ D.~R.~Yakovlev,$^{2,3}$   V.~Yu.~Ivanov,$^{4}$ D.~Kudlacik,$^{2}$  M.~V.~Eremin,$^{1}$ R.~V.~Pisarev,$^{3}$ and  M.~Bayer$^{2,5}$ } 
\affiliation{$^{1}$Institute of Physics, Kazan Federal University, 420008 Kazan, Russia}
\affiliation{$^{2}$Experimentelle Physik 2, Technische Universit\"at Dortmund, 44227 Dortmund, Germany}
\affiliation{$^{3}$Ioffe Institute, Russian Academy of Sciences, 194021 St. Petersburg, Russia}
\affiliation{$^{4}$Institute of Physics Polish Academy of Sciences, 02668 Warsaw, Poland}
\affiliation{$^{5}$Research Center FEMS, Technische Universit\"at Dortmund, 44227 Dortmund, Germany}

\date{\today}

\begin{abstract}
Nonreciprocity of light emission, when the radiation intensity differs for two opposite propagation directions, is a rare phenomenon in solids because it requires a violation of the crystal symmetry with respect to time-reversal. Such violation via time-reversal symmetry breaking can occur either due to an applied magnetic field or due to a magnetic ordering. We perform a detailed theoretical and experimental study of the photoluminescence (PL) nonreciprocity in the noncentrosymmetric tetragonal antiferromagnet CuB$_2$O$_4$, where this effect reaches 80\% below the N\'eel phase transition temperature of $T_N = 20$~K.  The effect is observed for three sets of extremely narrow exciton and exciton-magnon PL lines, associated with Frenkel excitons on the Cu$^{2+}$ ions in the magnetic $4b$ subsystem. A strong manifestation of the nonreciprocity of emission is found in certain geometries for the commensurate antiferromagnetic phase, as well as in other phases with incommensurate spin ordering. In accordance with the magnetic symmetry of CuB$_2$O$_4$, the nonreciprocity of emission is observed for light propagation along certain directions within the easy (001) plane. A rigorous quantum-mechanical analysis of the wave functions of the initial and final states of the Cu$^{2+}$ ions responsible for the PL is performed for various experimental geometries of the crystallographic axes and the applied magnetic field. The analysis confirms that the nonreciprocity of emission from Frenkel excitons in CuB$_2$O$_4$ is due to the interference of magnetic-dipole and electric-dipole transitions of antiferromagnetically ordered $4b$ spins of the Cu$^{2+}$ ions, in good agreement with the experimental data.
\end{abstract}
                              
\maketitle

\section{Introduction}
\label{SecI}

In optics, light reciprocity refers to a basic principle according to which the propagation, transmission or reflection of a light wave that follows the same path in forward and backward directions leads to the same result~\cite{Potton2004}.  Conversely, the principle of nonreciprocity applies to the situation where the properties of the light wave change when it travels in the opposite direction~\cite{Caloz2018}.  Both of these principles are applicable not only to optical phenomena, but also to the propagation of electromagnetic waves in general as well as to electron transport phenomena~\cite{Nagaosa2024}. 

The first experimental observation of optical nonreciprocity was the discovery of the Faraday effect, the rotation of the polarization plane of linearly polarized light that propagates in a material subject to an external magnetic field.  The main factor determining the occurrence of nonreciprocity in a medium is the violation of time invariance when the light direction is reversed. For reasons of symmetry, this effect can take place in any medium, in both paramagnets and diamagnets, however, only when an external magnetic field is applied.  In contrast, in magnetically-ordered media with spontaneously broken time reversal symmetry, such as ferromagnets and antiferromagnets, light nonreciprocity can be observed even without external magnetic field, provided that the sample is in a single-domain state. Symmetry aspects of nonreciprocity and propagation of electromagnetic radiation in magnetic media are discussed in Refs.~\onlinecite{Dionne2005,Szaller2013,Sheong2018,Caloz2018}.  

The principle of nonreciprocity applies to light emission, when the properties of local emitters in a medium demonstrate directionality. Nonreciprocity of emission  (NE) was found in magnetic ferroelectric (Ba,Sr)TiO$_3$ doped with Er$^{3+}$ ions~\cite{Shimada2006}, as well as in  La$_2$Ti$_2$O$_7$ doped with $R$ = Er, Eu and Nd, where it was termed as optical magnetoelectric effect~\cite{Shimada2007}. The microscopic origin of this effect is accounted for by the interference between electric-dipole and magnetic-dipole transitions via the spin-orbit interaction. However, the nonreciprocity of emission in paramagnets observed only in a magnetic field is rather small, not exceeding of 0.5\%. 

Much stronger effects are observed in magnetically-ordered crystals. Giant nonreciprocal emission reaching the contrast of 35\% was experimentally observed in the aniferromagnet CuB$_2$O$_4$~\cite{Toyoda2016}.  It was suggested that this effect, termed as direction-dependent luminescence (DDL), originates from the interference of the electric-dipole and magnetic-dipole transitions contributing to the emission of Cu$^{2+}$ ions from excited states to the ground state~\cite{Toyoda2016}. This interpretation was recently confirmed in microscopic theoretical analysis~\cite{Nurmukhametov2022}. However, the demand for a detailed theoretical analysis of the microscopic mechanisms responsible for the nonreciprocity of emission in CuB$_2$O$_4$ for various orientations of the antiferromagnetic spins and the magnetic field has remained unaccomplished.

Copper oxides offer a wide variety of physical properties that in many cases are associated with the possibility of the 3$d$ copper ion to have different valences, namely Cu$^{1+}$, Cu$^{2+}$ and Cu$^{3+}$. The Cu$^{1+}$ compounds with the closed 3$d^{10}$ orbital are diamagnetic. Compounds with partially-filled 3$d^9$ or 3$d^8$ orbitals are paramagnetic or magnetically ordered. High-$T_C$ superconductors are based on complex Cu$^{1+}$ and Cu$^{2+}$ oxides~\cite{Bednorz1986,AndersonScience1987,PickettRMP1989,DagottoRMP1994,DNBasov,Phillips2012,Sharma2021,SudipChakravarty2001,Zhang,Sachdev2003}.  The chemically simplest Cu$^{2+}$ oxide CuO is an antiferromagnet and multiferroic~\cite{CuOKimura2022}. From the point of view  of electrical properties, this material is a semiconductor, which finds application, e.g., as a photoresistor~\cite{Dolai2017}.  

However, most mixed-valency cuprates are non-superconducting. For example, lithium cuprate Li$_{2}$Cu$^{(1+/2+)}$O$_2$ is used for a wide range of applications due to high lithium diffusion through the layered structure~\cite{Palacios-Romero2023}. The quasi-one-dimensional mixed-valency antiferromagnets LiCu$^{1+}$Cu$^{2+}$O$_2$ and LiCu$^{1+}$Cu$_2^{2+}$O$_3$ are ionic conductors~\cite{Hibble1990,Berger1992,Zatsepin}. From the point of view of their magnetic properties, these lithium cuprates demonstrate complex spin ordering~\cite{Capogna2005,Drechsler2005,Gippius2008,Svistov2010} and Zhang-Rice singlets~\cite{PRL2024}. 

The potassium Cu$^{2+}$ fluorides K$_2$CuF$_4$ and KCuF$_3$ are well known as prominent representatives of materials showing the cooperative Jahn-Teller effect~\cite{Khomskii1973,Kugel1973}. New types of collective spin excitations (spinons) have been detected in K$_2$CuSO$_4$Br$_2$ using the magnetic resonance technique~\cite{Smirnov2025}. It is interesting to note that for many centuries, and even millennia, copper oxide compounds have been widely used as green and blue pigments~\cite{Svarcova2021}. These particular colors are determined by optical absorption in the visible spectral range related to transitions between the ground and excited 3d$^9$ states of the Cu$^{2+}$ ions split by the local crystal field~\cite{Burns1993}. 

In this paper, we study experimentally and theoretically the nonreciprocity of emission in the copper metaborate CuB$_2$O$_4$, which in many aspects is radically different from other compounds with partially filled 3$d^{n}$ and 4$f^{n}$ magnetic ions. It has a complex magnetic phase diagram with commensurate and incommensurate antiferromagnetic ordering~\cite{Pankrats2018,Lai2024}.
Very bright photoluminescence related to Cu$^{2+}$ ions was observed in~\cite{Kudlacik_2020}. 
This material is characterized by a strong antiferromagnetic dichroism~\cite{Boldyrev2015} and a nonreciprocal second harmonic generation~\cite{Mund2021,Toyoda2021}.

We experimentally examine various geometries defined by the mutual orientation of the emission wave vector, the crystallographic axes, and the external magnetic field. Experiments are performed in different magnetic phases below and above the N\'eel temperature of $T_N = 20$~K selected by varying the sample temperature as well as the magnetic field orientation and its strength. For these geometries the microscopic theoretical analysis and model calculations of the emission-directionality diagrams are performed. Good agreement is found between the experimental results and the theoretical predictions.
 
The paper is organized as follows. The crystal and magnetic structures of CuB$_2$O$_4$ are discussed in Sec.~\ref{Structure}. In Section~\ref{sec:Theory} the microscopic theory is presented giving the model results for the nonreciprocity of emission in the main experimental geometries. Experimental details are given in Sec.~\ref{sec:Experimentals}. Section~\ref{Sec:Results} presents the experimental data. Here, we demonstrate all geometries at $T = 15$~K. For a particular geometry, we present the data at $T = 1.7$, 5 and 23~K corresponding to various magnetic phases. In Section~\ref{Sec:Discussion} we discuss the experimental results and compare them with the theoretical predictions.


\section{Crystal and magnetic structure of C\lowercase{u}B$_2$O$_4$ and exciton transitions}
\label{Structure}

\subsection{Crystal structure}
\label{SubSec:1A}

The structure of CuB$_{2}$O$_{4}$ is described by the point group -42$m$ and the space group $I$-42$d$~\cite{Martinez1971,Abdullaev1981}. The crystallographic unit cell contains four Cu$^{2+}$ ions at the $4b$ positions and the eight ions at the $8d$ positions. The transition from paramagnetic to  antiferromagnetic state due to the $4b$ subsystem takes place at  $T_N=20$~K. Magnetic ordering of the $8d$ subsystem occurs only below 10~K, but this subsystem does not contribute to the photoluminescence and we do not discuss it in this paper.

\begin{figure}[hbt]
\includegraphics[width=\linewidth]{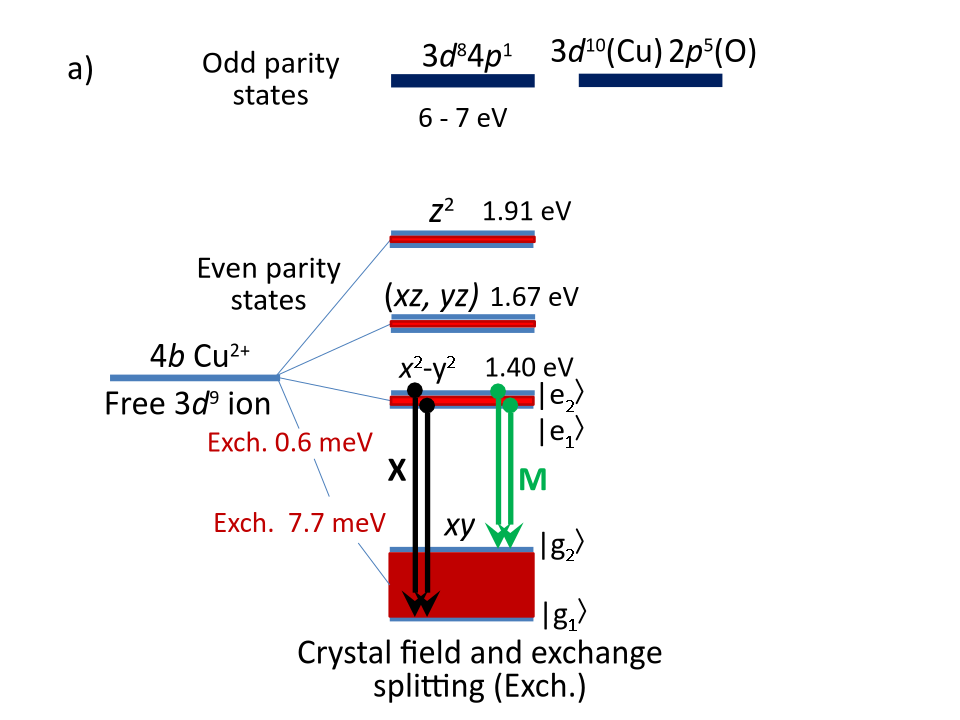}
\includegraphics[width=\linewidth]{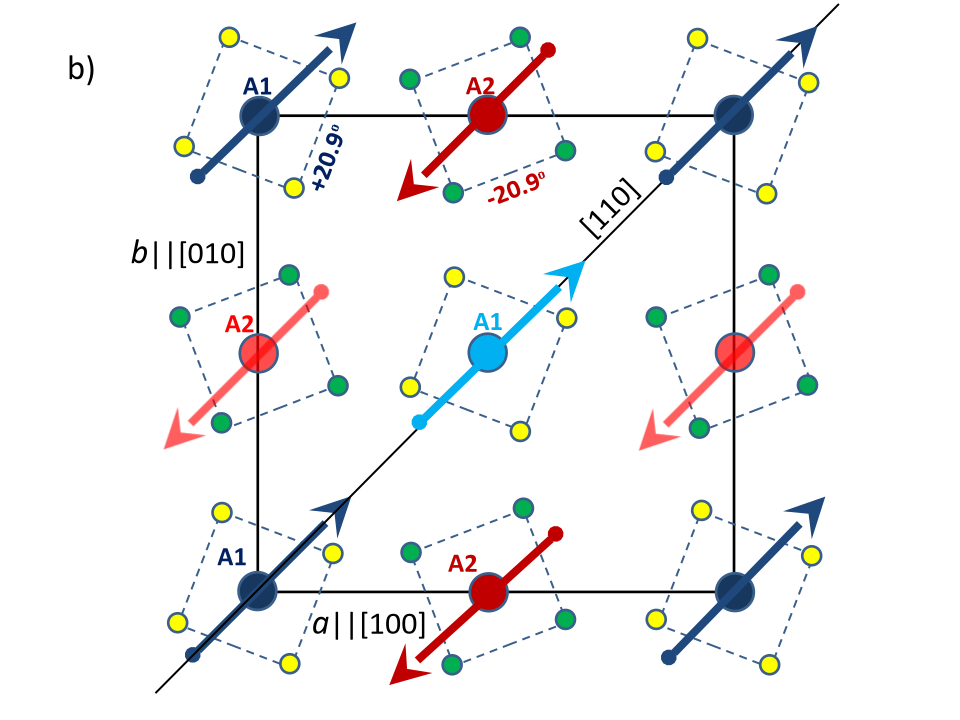}
   \caption{(a) Crystal field and exchange splitting of exciton states in the $4b$ subsystem of the Cu$^{2+}$ ions. The even parity of the $3d^9$ states is broken by the perturbation from the $3d^84p^1$ odd states.
(b) Magnetic structure of the $4b$ subsystem with spins antiferromagnetically ordered within the (001) easy plane. Four types of antiferromagnetic domains can exist in CuB$_{2}$O$_{4}$ with the spins oriented along the [110], [-1-10], [-110] and [1-10] easy axes. The theoretical analysis is made for the single domain state [110]. Note that the A1 and A2 planar oxygen squares surrounding the $4b$ Cu$^{2+}$ ions are rotated by $\pm$24$^\circ$ in opposite directions.
   }
\label{fig:fig1}     
 \end{figure}

According to the structural analysis, the four Cu$^{2+}$ ions at the 4$b$ positions have the same local symmetry, but are divided into pairs of A1 and A2 subsystems, as shown in Fig.~\ref{fig:fig1}(a). These subsystems of identical CuO$_4$ squares are turned by $\pm$24$^\circ$ in opposite directions in the ($ab$)-plane. This structural feature is important for developing a basic microscopic theory.

\subsection{Magnetic structure}
\label{SubSec:2A}

We note that in the commensurate magnetic phase between $T=9$~K and $T_N=20$~K~\cite{Boehm2003} at equilibrium, the Cu$^{2+}$ spins of the 4$b$ subsystem are oriented along the [110]-type axis within the mirror plane $m$ and this structure with the symmetry -42$m$' is altermagnetic~\cite{Radaelli2024}.   

As demonstrated in Refs.~\onlinecite{Pankrats2018,Lai2024}, the magnetic phase diagram of CuB$_{2}$O$_{4}$ is strongly dependent on the applied magnetic field strength and its orientation with respect to the crystallographic axes. In our theoretical analysis, we will discuss the case when the spins are in the collinear phase and no spin-flop transitions occur~\cite{Pankrats2018}. Two main geometries for the magnetic field orientation must be distinguished. The first one for  the field oriented within the easy ($ab$)-plane in which the magnetocrystalline anisotropy is small.  
The second one for the field applied perpendicular to the ($ab$)-plane along the $c$-axis, or $\mathbf{B}\parallel$[001]. 


Typically, in the absence of an external magnetic field, the sample is in a multi-domain state, and the results can vary depending on the relative fractions of the different domains. By symmetry, in zero field four types of domains can exist, namely with the antiferromagnetic vector $\mathbf{L} \parallel [110]$ or [-1-10], and $\mathbf{L}\parallel$[-110] or [1-10]. In the extreme case, when the domains with opposite spin (or opposite $\mathbf{L}$ vector) are equally represented, their contributions to the nonreciprocal effects cancel out each other. In order to observe the nonreciprocal effects in experiment, one has to establish a single-domain state, for example, by applying a  magnetic field in a proper direction which is sufficient to orient the magnetic domains. In the theoretical analysis, we assume that the sample is in a single-domain state with the vector $\mathbf{L}\parallel$[110] so that it is oriented along an easy axis.

\subsection{Excitons}
\label{SubSec:3A}

The optical properties of CuB$_2$O$_4$ in the near-infrared, visible, and ultraviolet spectral range are determined by absorption and emission from the electronic states of the 4$b$ and 8$d$ Cu$^{2+}$ ions~\cite{Pisarev2011,Kudlacik_2020}. The optical exciton transitions of the 4$b$ copper ions are positioned at 1.405, 1.667, and 1.902~eV, as shown in Fig.~1(b) of Ref.~\onlinecite{Pisarev2011}. The lowest energy absorption band in corresponding spectra starts with the narrow zero-phonon line at ~1.405~eV (at $T=1.6$~K) which is related to the 4$b$ subsystem. Photoluminescence studies allowed the observation of a rich spectrum below the 1.405~eV exciton line that spans over a broad spectral range of exciton, exciton-phonon, and exciton-magnon lines~\cite{Kudlacik_2020}. The theoretical analysis presented in the Sec.~\ref{sec:Theory} focuses on the PL emission from the lowest-in-energy exciton at 1.405~eV, related to the 4$b$ positions.


\section{Theory of the nonreciprocity of emission in C\lowercase{u}B$_2$O$_4$}
\label{sec:Theory}
 
We develop a microscopic theory in which we model the electronic structure of both the A1 and A2 types of the magnetic Cu$^{2+}$ ions using a single-ion approach. First, we determine the energy levels and the wave functions of the Cu$^{2+}$ ions at the 4$b$ positions. We then use these wave functions to compute the matrix elements of the electric-dipole (ED) and magnetic-dipole (MD) transitions. Both of these mechanisms contribute to the net transition probability and, consequently, to the PL intensity and the nonreciprocity of emission.

A microscopic theory of the nonreciprocity of emission was previously developed in Ref.~\onlinecite{Nurmukhametov2022}. Here, we extend this theory by modifying the previous model and revising the method for calculating transition probabilities. Specifically, we incorporate the exchange field operator that acts on the first excited state at 1.405 eV at the A1 and A2 positions, as identified in Ref.~\onlinecite{Kopteva_2022}. Then, we carry out calculations in the system of tetragonal crystallographic axes rather than in the system of local axes for each of the A1 and A2 sets of the Cu$^{2+}$ $4b$ ions.
 
\subsection{Energy levels and wave functions of Cu$^{2+}$ ions}
\label{SubSec:2A}

The energy levels and wave functions of the Cu$^{2+}$ ions at the  A1 and A2 positions are calculated by diagonalizing the Hamiltonian:
\begin{equation}
\label{eq:hamiltonian}
 H = H_{cf} + H_{so} + H_{ex}^{gg} + H_{ex}^{ge} + H_{Z},
\end{equation}
Here, $H_{cf}$, $H_{so}$, and $H_{Z}$ denote the standard crystal field acting on the Cu$^{2+}$ ions, the spin-orbit interaction, and the external magnetic field operators, respectively. The crystal field parameters $B_q^{(k)}$ were determined in Ref.~\onlinecite{Eremin2021} using experimental data from absorption~\cite{Pisarev2011}.
\begin{equation}
\label{eq:h_cf_so_z}
\begin{aligned}
& H_{cf} = \sum B_q^{(k)}C_q^{(k)},\\ & H_{so} = \lambda \mathbf{ls}, \\ & H_{Z} = \mu_{B} \mathbf{B} (\mathbf{l} + 2 \mathbf{s}).
 \end{aligned}
\end{equation}

The terms $H_{ex}^{gg}$ and $H_{ex}^{ge}$ in Eq.~\eqref{eq:hamiltonian} represent the exchange field operators acting on the ground and excited states of the copper ions. A key distinction of this approach from the approaches used in previous studies~\cite{Eremin2021, Nurmukhametov2022, Boldyrev2023} is the choice of the basis for the copper ion states. Here, we employ orbital wave functions defined in the crystallographic coordinate system $a\parallel[100]$, $b\parallel[010]$, $c\parallel[001]$: $| \epsilon \rangle  = | x^2-y^2 \rangle, \; | \zeta \rangle = | xy \rangle, \; |\theta \rangle  = | 3z^2-r^2 \rangle, \; | \eta \rangle = | xz \rangle, \; | \xi \rangle = | yz \rangle$, along with the spin states $| \pm 1/2 \rangle$.  This basis is particularly suitable for analyzing variations of the exchange interaction parameters in the ground and excited states as a function of the direction of the external magnetic field, especially for the excited doublet responsible for the PL emission and nonreciprocity. 

According to the semi-empirical Goodenough-Kanamori rules, the strength and sign of the exchange interaction parameter are highly sensitive to the wave functions~\cite{Goodenough1955, Kanamori1959}. As discussed in Ref.~\onlinecite{Toyoda2015}, the influence of the magnetic field on the excited states must account for their mixing  with the ground-state wave function $| \epsilon \rangle  = | x^2-y^2 \rangle$ due to the spin-orbit coupling. Consequently, variations in the orbitally selective contributions to the exchange fields acting on the ground and excited states must be self-consistently matched. With this in mind, we express the exchange interaction operators in Eq.~\eqref{eq:hamiltonian} as follows:
\begin{equation}
\label{eq:h_ex}
\begin{aligned}
& H_{ex}^{gg} = \frac{J_{gg}}{2} \left( |g'\rangle \langle g'| \sum_{i=1}^{4} \langle \mathbf{S}_{cu} \rangle \mathbf{S} + \sum_{i=1}^{4} \langle \mathbf{S}_{cu} \rangle \mathbf{S} |g'\rangle \langle g'| \right), \\
& H_{ex}^{ge} = \frac{J_{ge}}{2} \left( |e'\rangle \langle e'| \sum_{i=1}^{4} \langle \mathbf{S'}_{cu} \rangle \mathbf{S} + \sum_{i=1}^{4} \langle \mathbf{S'}_{cu} \rangle \mathbf{S} |e'\rangle \langle e'| \right) \,.
\end{aligned}
\end{equation}
Here $|g'\rangle, |e'\rangle$ are the wave functions of the ground and first excited doublets at ~1.4 eV energy obtained by diagonalizing the Hamiltonian $ H_{cf} + H_{so} $; $|g'\rangle \langle g'|$ and $|e'\rangle \langle e'|$ are the corresponding projection operators; $J_{gg}$ and $J_{ge}$ are the parameters of exchange interaction. $\langle \mathbf{S}_{cu} \rangle$ and  $\langle \mathbf{S'}_{cu} \rangle$ denote the average spins of neighboring copper ions when the ion in question is in its ground and excited states, respectively. Below, the antiferromagnetic vector $\mathbf{L} \sim \mathbf{S}_{A1} - \mathbf{S}_{A2}$ is used to describe the magnetic structure. It can be easily seen that $\mathbf{L} \parallel \langle \mathbf{S}_{cu} \rangle$, hence when we say, that the external magnetic field rotates the antiferromagnetic vector, it means that the average spin directions also rotate, which affects the emission intensity. To characterize the magnetic structure, we employ the antiferromagnetic order parameter $\mathbf{L} \sim \mathbf{S}_{A1} - \mathbf{S}_{A2}$. Given that $\mathbf{L}$ is parallel to $\langle \mathbf{S}_{cu} \rangle$ in the commensurate phase, the rotation of $\mathbf{L}$ when applying an external magnetic field is equivalent with the rotation of the average spin moments.

Note that the representations of the wave functions can be found relatively straightforward for the A1 and A2 sites. For example, in the local coordinate system, the ground and excited states involved in light emission are written in the crystal field as $| g_{l} \rangle = | \epsilon \rangle$ and $|e_{l} \rangle = |\zeta \rangle$, respectively. Then, in the crystallographic coordinate system, we get: $|g_{\rm A1} \rangle = 0.6664 | \epsilon \rangle + 0.7456 | \zeta \rangle$, $|e_{\rm A1} \rangle = -0.7456 | \epsilon \rangle + 0.6664 | \zeta \rangle$, $|g_{\rm A2} \rangle = -0.6664 | \epsilon \rangle + 0.7456 | \zeta \rangle$, and $|e_{\rm A2} \rangle = 0.7456 | \epsilon \rangle + 0.6664 | \zeta \rangle$.

The crystal field parameters $B_{q}^{(k)}$ calculated in the crystallographic coordinate system \cite{Eremin2021} are expressed in eV units as follows:
\begin{equation}
\label{eq:cf_params}
\begin{aligned}
 &B^{(2)}_{0} = -2.18, \,\, B^{(4)}_{0} = 1.17, \,\, \\ 
 &B^{(4)}_{4}(A1) = 0.20 + 1.75i, \\
 &B^{(4)}_{4}(A2) = 0.20 - 1.75i, \\
 \end{aligned}
\end{equation}

The parameter of the exchange interaction between the two ground substates of the Cu$^{2+}$ ions is taken as $J_{gg} = 3.62$~meV~\cite{Boehm2002}. The exchange parameter between the excited state $| e' \rangle$ and the ground state $| g' \rangle$ of the nearest-neighbor copper ions was recently determined as $J_{ge} = -0.3$~meV in Ref.~\onlinecite{Kopteva_2022}. The spin-orbit interaction parameter is $\lambda = -0.1$~eV~\cite{AbragamBleaney}. The Zeeman term $H_Z$ is on the order of several wave numbers in cm$^{-1}$, and, therefore, is small compared with the exchange and spin-orbit interactions, see Eq.~\eqref{eq:hamiltonian}. The primary effect of an external magnetic field is to control the spin orientation. This realignment changes the exchange interaction and, consequently, the resulting electronic wavefunctions.

As demonstrated in Ref.~\onlinecite{Pankrats2018}, the magnetic phase diagram is strongly dependent on the magnetic field strength and its orientation with respect to the crystallographic axes. In our theoretical analysis, we will discuss the case when the spins are in the collinear phase and no spin-flop transitions occur~\cite{Pankrats2018}. Two main geometries for the magnetic field orientation must be distinguished. The first one for the field oriented within the easy ($ab$)-plane in which the magnetocrystalline anisotropy is small. Here in sufficiently strong field on the order of 0.2~T the antiferromagnetic vector $\mathbf{L}$ is oriented perpendicular to the direction of the applied field. For the field within the $ab$-plane, two important orientations are $\mathbf{B}\parallel$[100] or [010], and $\mathbf{B}\parallel$[110] or [-110]. The second important orientation is for the field applied perpendicular to the ($ab$)-plane along the $c$-axis, or $\mathbf{B}\parallel$[001].

\subsection{Interaction of electrons with electric field of light wave}
\label{SubSec:2B}

The effective operator of the interaction energy of a $3d$-electron with the electric field of the light wave is written in the form:
\begin{equation}
\label{eq:HE}
H_{E}=\sum_{\substack{k=2,4 \\ p = 1,3,5 \\ t = \pm 2}}\left\{E^{(1)} U^{(k)}\right\}_{t}^{(p)} D_{t}^{(1 k) p}.
\end{equation}
Here, $D_{t}^{(1k)p^1}$ are the interaction parameters which include two types of terms. The first type is due to the mixing of the main configuration of the even states $3d^9$ with the odd excited states $3d^{8}4p^1$ by the crystalline field, as in the theory of forced electric-dipole transitions. The second type is due to the mixing with charge transfer transitions from oxygen to the 4$b$ copper ions~\cite{Eremin2019JETP}. The round parentheses give the direct product of the spherical components of the electric field vector $E^{(1)}$ and the unitary irreducible tensor operator $U^{(k)}$ ($\hat{U}_{q^\prime}^{(k)}$ are reduced for the sake of brevity):
\begin{equation}
\label{eq:electricalProduct}
\begin{aligned}
&\left\{E^{(1)} \hat{U}^{(k)}\right\}_{t}^{(p)} = \\
& \sqrt{2p+1} \sum_{q,q^\prime} (-1)^{1-k+t} 
\begin{pmatrix}
1 & k & p \\
q & q^\prime & -t
\end{pmatrix}
E_{q}^{(1)} \hat{U}_{q^\prime}^{(k)},
\end{aligned}
\end{equation}
\begin{equation}
\label{eq:electricalComponents}
E_{0}^{(1)} = E_z; \; E_{\pm 1}^{(1)} = \mp \frac{1}{\sqrt{2}} (E_x \pm i E_y).
\end{equation}

Additional selection rules are imposed by the point group symmetry inversion operation S4 of the copper positions in the 4$b$ subsystem of antiferromagnetically coupled spins. Ultimately, it is easy to establish that the following values are nonzero: $D_{\mp2}^{(12)3}$, $D_{\mp2}^{(14)3}$ and $D_{\mp2}^{(14)5}$, which have the dimension of an electric-dipole moment. The matrix elements of operator $U^{(k)}$ are calculated using the 3-$j$ symbols:
\begin{equation}
\label{eq:MatrixU}
\langle dm_l \lvert \hat{U}_{q^\prime}^{(k)} \rvert dm_l^\prime \rangle = (-1)^{2-m_l} 
\begin{pmatrix}
2 & k & 2 \\
-m_l & q^\prime & m_l^\prime
\end{pmatrix}.
\end{equation}

The $D_{t}^{(1k)p}$ values are calculated in the local coordinate systems of the A1 and A2 ions and then translated to the crystallographic system as in Ref.~\onlinecite{Boldyrev2023}. Then they are rotated at an angle $\phi=24^{\circ}$ using the following relationship:
\begin{equation}
\label{eq:DkptLocal}
{D^\prime}_t^{(1k)p} = D_t^{(1k)p} \exp(-it\phi).
\end{equation}
The parameter values for the A1 and A2 positions are related to each other through the following equations~\cite{Boldyrev2023}:
\begin{equation}
\label{eq:DkptA1ToA2}
\begin{aligned}
& D_2^{(12) 3}\left(\mathrm{A}_1\right)=-D_2^{(12) 3^*}\left(\mathrm{A}_2\right), \\
& D_2^{(14) 3}\left(\mathrm{A}_1\right)=-D_2^{(14) 3^*}\left(\mathrm{A}_2\right), \\
& D_2^{(14) 5}\left(\mathrm{A}_1\right)=-D_2^{(14) 5^*}\left(\mathrm{A}_2\right) .
\end{aligned}
\end{equation}

\subsection{Angular dependence of emission intensity in external magnetic field}
\label{SubSec:2C}

The intensity of the transitions between the excited $| e \rangle$ and the ground $| g \rangle$ states is proportional to the sum of the squares of the moduli of the matrix elements across the 4$b$ positions of the copper ions in the unit cell:
\begin{equation}
\label{eq:Ieg}
I_{eg} \sim \frac{1}{Z} \sum_{i} \left| \langle e_i | H_E + H_M | g \rangle \right|^{2} e^{- \frac{E_i}{k_B T}}.
\end{equation}
Here $Z$ is the partition function, $H_M$ is the standard operator of magnetic-dipole (MD) transitions, $H_E$ is the operator of electric-dipole (ED) transitions, as introduced in Eq.~\eqref{eq:HE}, $E_i$ is the energy of the $e_i$ state.

According to the general theory of radiation, it is natural to assume that for the preferential emission direction the transition probability between the initial and final states is the largest. In this regard, the nonreciprocity of emission and the  nonreciprocity of absorption~\cite{Toyoda2015,Toyoda2019} are two different optical phenomena, which have a common origin. Often it is difficult to separate these two phenomena experimentally. Especially, when emission and absorption occur at the same energy, as the properties of the emitted light are modified during its passage through the crystal. This difficulty can be avoided in the case of Stokes emission lines, which are shifted to lower energy from the absorption line. In fact, this is the case realized in CuB$_2$O$_4$~\cite{Kudlacik_2020}. 

The probabilities of electric-dipole transitions for a free copper ion are several orders of magnitude larger than those of magnetic-dipole transitions. However, within the states of the $3d^9$ configuration, electric-dipole transitions are forbidden by parity. The 3$d^9$ transitions observed in experiment appear due to weak admixture of states of opposite parity. In our case these are $3d^{8}4p^1$ states with charge transfer from copper to oxygen. As a result, the ED and MD contributions have comparable magnitude and  interference between them becomes possible and effective.

\begin{figure*}[hbt]
  \includegraphics[width=\linewidth]{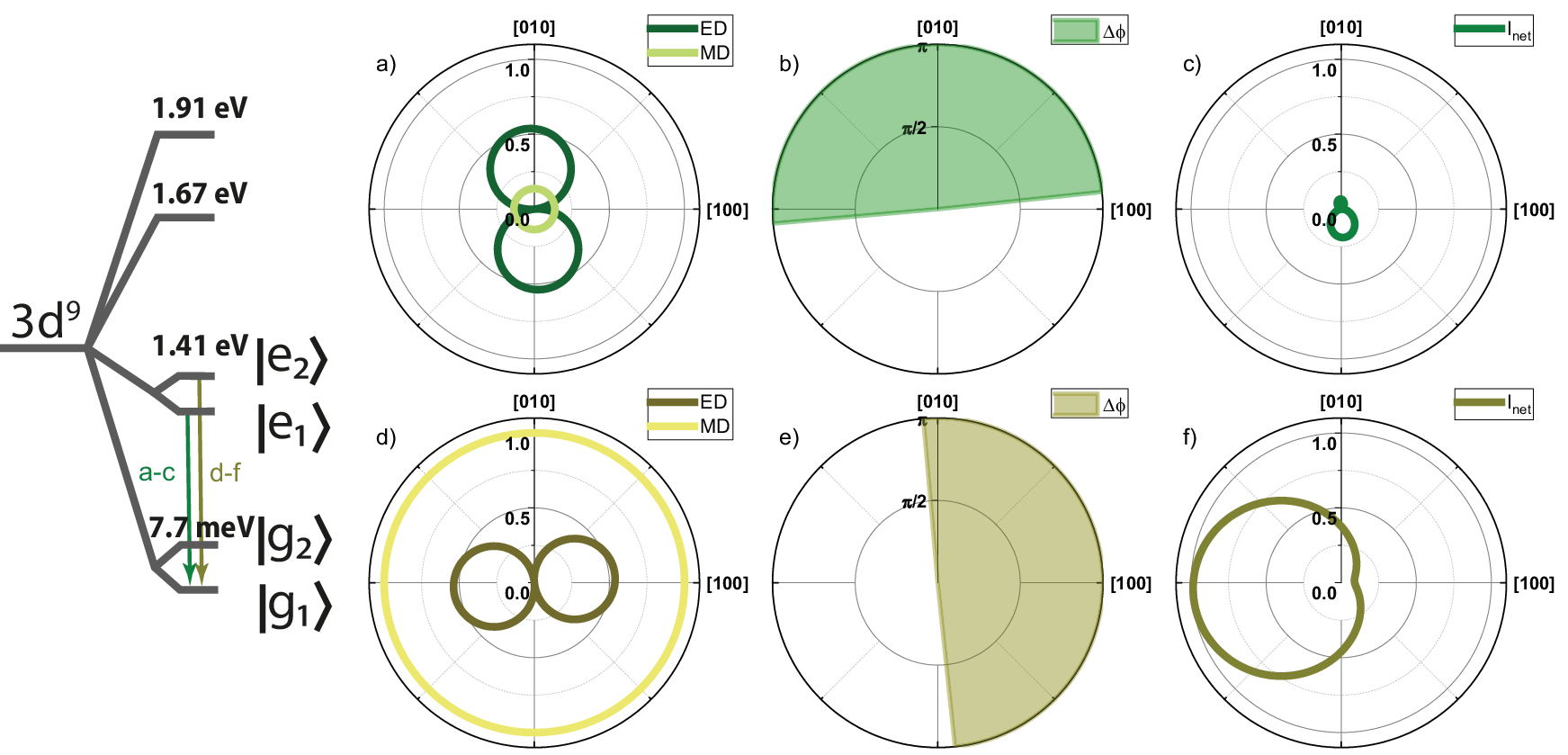}
   \caption{Illustration of the interference between the ED and MD transitions from the (a-c) lower and (d-f) upper components of the first excited doublet to the ground state. The calculations were performed for the A1 copper site under an external magnetic field of $B=0.5$~T rotating in the crystal \textit{(ab)}-plane at $T=15$~K. The emitted light wave is configured with $\mathbf{k} \parallel a$, $\mathbf{E^\omega} \parallel b$, and $\mathbf{H^\omega} \parallel c$. The antiferromagnetic vector $\mathbf{L}$ is constrained to being perpendicular to $\mathbf{B}$. The polar angle represents the orientation of $\mathbf{B}$ relative to the \textit{a} [100] crystal axis. The radial values represent: (a,d) the normalized moduli of the ED and MD matrix elements, (b,e) the phase difference between them, and (c,f) the resulting net transition intensity. Depending on the phase relationship between the ED and MD contributions, the net transition intensities (c,f) exhibit enhancement (constructive interference) or suppression (destructive interference).
   }
\label{fig:interfereceContributuons}     
 \end{figure*}

Before proceeding to the detailed modeling, we first illustrate the interference that occurs in the transitions from the first excited doublet to the ground state of the $4b$ Cu$^{2+}$ ion. As shown in Fig.~\ref{fig:interfereceContributuons} for the A1 site, the matrix elements of ED and MD transitions are comparable (a,d), with their relative phase difference changing from 0 to $\pi$/2 and then to $\pi$, depending on the external magnetic field direction in the ($ab$)-plane (b,e). When the matrix elements are in phase, constructive interference enhances the net intensity, leading to stronger emission for this configuration. Conversely, when they are out of phase, the contributions partly cancel out resulting in weaker emission (c,f). 

\begin{figure*}[hbt]
\begin{center}
  \includegraphics[width=\linewidth ]{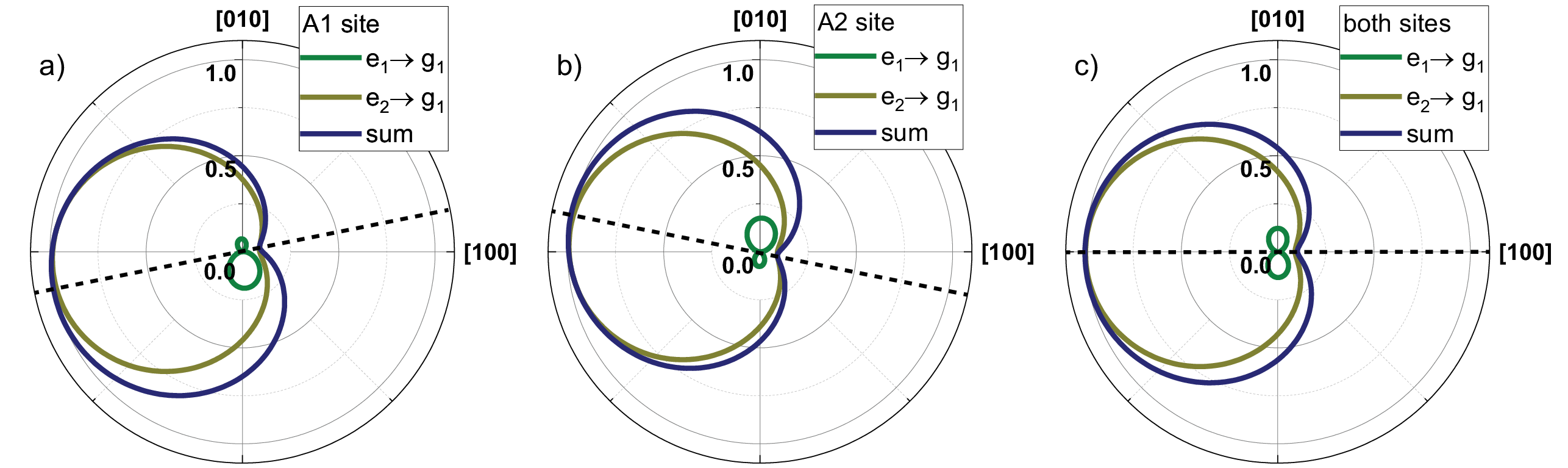}
\end{center}
  \caption{Calculated angular dependence of the emission intensities from the components of the first excited doublet to the ground state calculated for the (a) A1 position, (b) A2 position, and (c) sum of both positions under an external magnetic field of $B=0.5$~T rotating in the crystal \textit{(ab)}-plane at $T=15$~K. The emitted light wave is configured with $\mathbf{k} {\parallel} a$, $\mathbf{E^\omega} {\parallel} b$, and $\mathbf{H^\omega} {\parallel} c$. The antiferromagnetic vector $\mathbf{L}$ is constrained to be perpendicular to $\mathbf{B}$. 
  The polar angle represents the orientation of $\mathbf{B}$ relative to the \textit{a} [100] crystal axis. 
  The radial values reflect the normalized intensities calculated with Eq.~\eqref{eq:Ieg}.
  Note that while diagrams for singular A1 and A2 sites are tilted, the resulting net intensity is symmetric with respect to the \textit{a}-axis.
   }
   \label{fig:interfereceBothSites}  
 \end{figure*}

Note that the patterns in the diagrams of Fig.~\ref{fig:interfereceContributuons}  are slightly tilted with respect to the crystallographic axes because of the local symmetry of the A1 site under consideration. For the A2 site, the tilting is reversed (compare Figs.~\ref{fig:interfereceBothSites}(a) and \ref{fig:interfereceBothSites}(b)), so that the net emission diagram summed over all $4b$ copper ions is aligned along the crystal axis [100], as we show in Fig.~\ref{fig:interfereceBothSites}(c).

\begin{figure*}[hbt]
\begin{center}
  \includegraphics[width=\linewidth ]{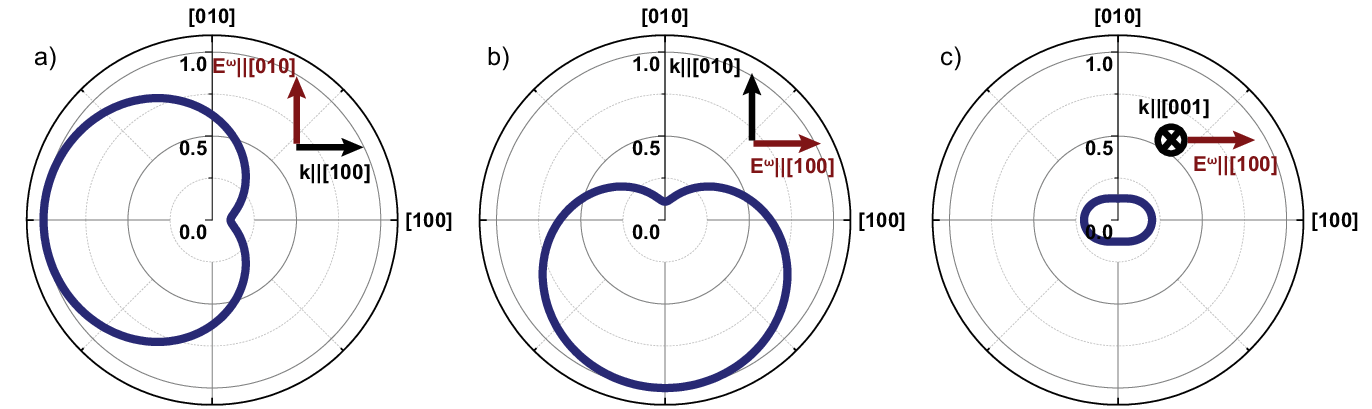}
\end{center}
  \caption{Calculated angular dependence of the emission intensities from the components of the first excited doublet to the ground state under an external magnetic field of $B=0.5$~T rotating in the crystal \textit{(ab)}-plane at $T=15$~K. The emitted light wave geometries for each diagram are shown in the insets. The antiferromagnetic vector $\mathbf{L}$ is constrained to be perpendicular to $\mathbf{B}$.  The polar angle represents the orientation of $\mathbf{B}$ relative to the \textit{a} [100] crystal axis. The radial values reflect the normalized intensities calculated with Eq.~\eqref{eq:Ieg}. In panel (a), the maximum nonreciprocity occurs when $\mathbf{B} {\parallel} a$, whereas in panel (b), it occurs when $\mathbf{B} {\parallel} b$, consistent with the experimental observations. Panel (c) shows the absence of nonreciprocity for $\mathbf{k} {\parallel} c$, similar results appear for fields rotating in the \textit{(ac)}- and \textit{(bc)}-plane with $\mathbf{k} {\parallel} c$.   }
   \label{fig:fieldDDL}
 \end{figure*}

\begin{figure}[hbt]
\begin{center}
  \includegraphics[width=\linewidth ]{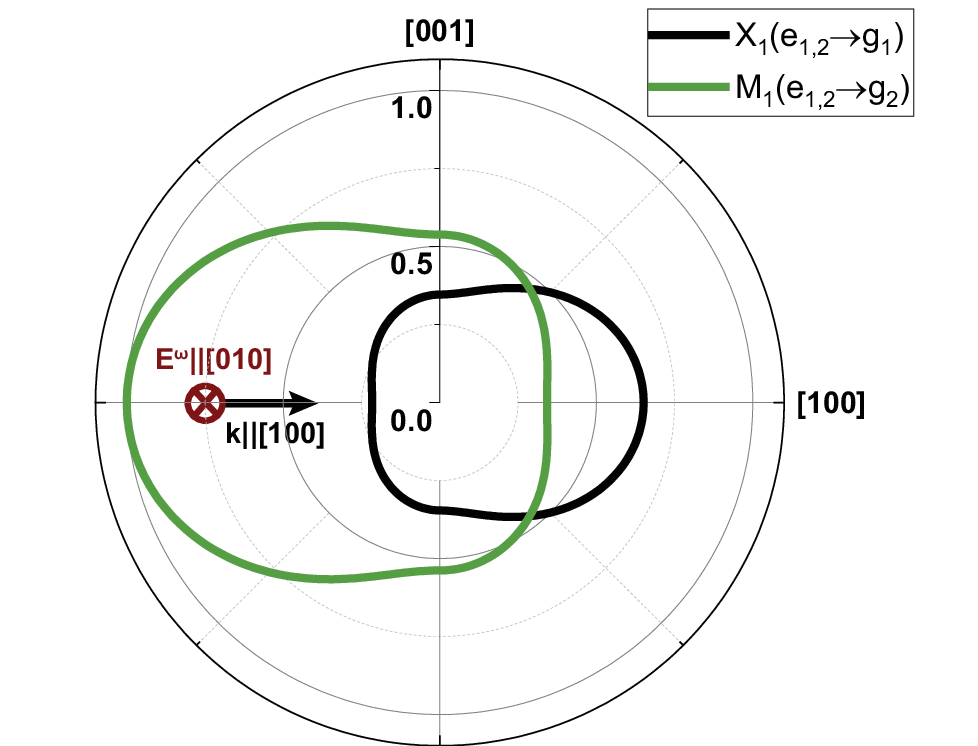}
\end{center}
  \caption{ Calculated angular dependence of the emission intensities from the components of the first excited doublet to the lower (X1) and upper (M1) components of the ground-state doublet.  The calculations were performed for an external magnetic field of $B=0.5$~T rotating in the crystal \textit{(ac)}-plane at $T=15$~K. The emitted light wave is configured with $\mathbf{k} {\parallel} a$, $\mathbf{E^\omega} {\parallel} b$, and $\mathbf{H^\omega} {\parallel} c$.
    The polar angle $\phi$ represents the orientation of $\mathbf{B}$ relative to the \textit{a} [100] crystal axis.
    The radial values reflect the normalized intensities calculated with Eq.~\eqref{eq:Ieg}.
    The antiferromagnetic vector is fixed along $\mathbf{L} {\parallel} b$. 
    For $\mathbf{B} {\parallel} a$ ($\phi = 0^\circ$), the crystal is treated as a single domain. As the field rotates away from the \textit{a}-axis, the population of domains with $\mathbf{L}$ and $-\mathbf{L}$ is assumed to vary proportionally to $\sin^2(\phi)$ achieving equal amounts of $\mathbf{L}$ and $-\mathbf{L}$ domains at $\mathbf{B} \parallel c$ ($\phi = 90^\circ$). While this simple model provides a sufficient approximation to reproduce the experimental data, a precise calculation of the relationship between $\mathbf{B}$ and $\mathbf{L}$ for fields close to the \textit{c}-axis is complex and is beyond the scope of this work. The same diagram appears for fields in the \textit{(bc)}-plane and the emitted light wave $\mathbf{k} {\parallel} b$, $\mathbf{E^\omega} {\parallel} a$, and $\mathbf{H^\omega} {\parallel} c$. 
       }
   \label{fig:ddlX1M1}  
 \end{figure}

The calculated diagrams of the relative emission intensity upon rotation of the external magnetic field $B=0.5$~T are shown in Figs.~\ref{fig:fieldDDL} and \ref{fig:ddlX1M1}. Note that an increase of the magnetic field strength up to 10 T results in only minor modifications to the diagrams. This shows that the role of the Zeeman term is weak. The principal effects of the external magnetic field are in preparing monodomain sample, in establishing the commensurate phase, and in tuning the exchange interaction by rotating $\mathbf{L}$. 

The emission nonreciprocity for the transition to the upper component of the ground doublet (M1 experimental line, Fig.~\ref{fig:fig1} green arrows) is opposite to the one for the lower component (X1 experimental line, Fig.~\ref{fig:fig1} black arrows), as shown in Fig.~\ref{fig:ddlX1M1}. 

As one can see in Figs.~\ref{fig:fieldDDL}(a) and \ref{fig:fieldDDL}(b), the maximum effect of nonreciprocity of emission is expected for the geometries $k_aB_a$ and $k_bB_b$. For $\textbf{k} \parallel c$ (Fig.~\ref{fig:fieldDDL}(c)) the nonreciprocity effect is absent.

\begin{figure*}[hbt]
  \includegraphics[width=\linewidth]{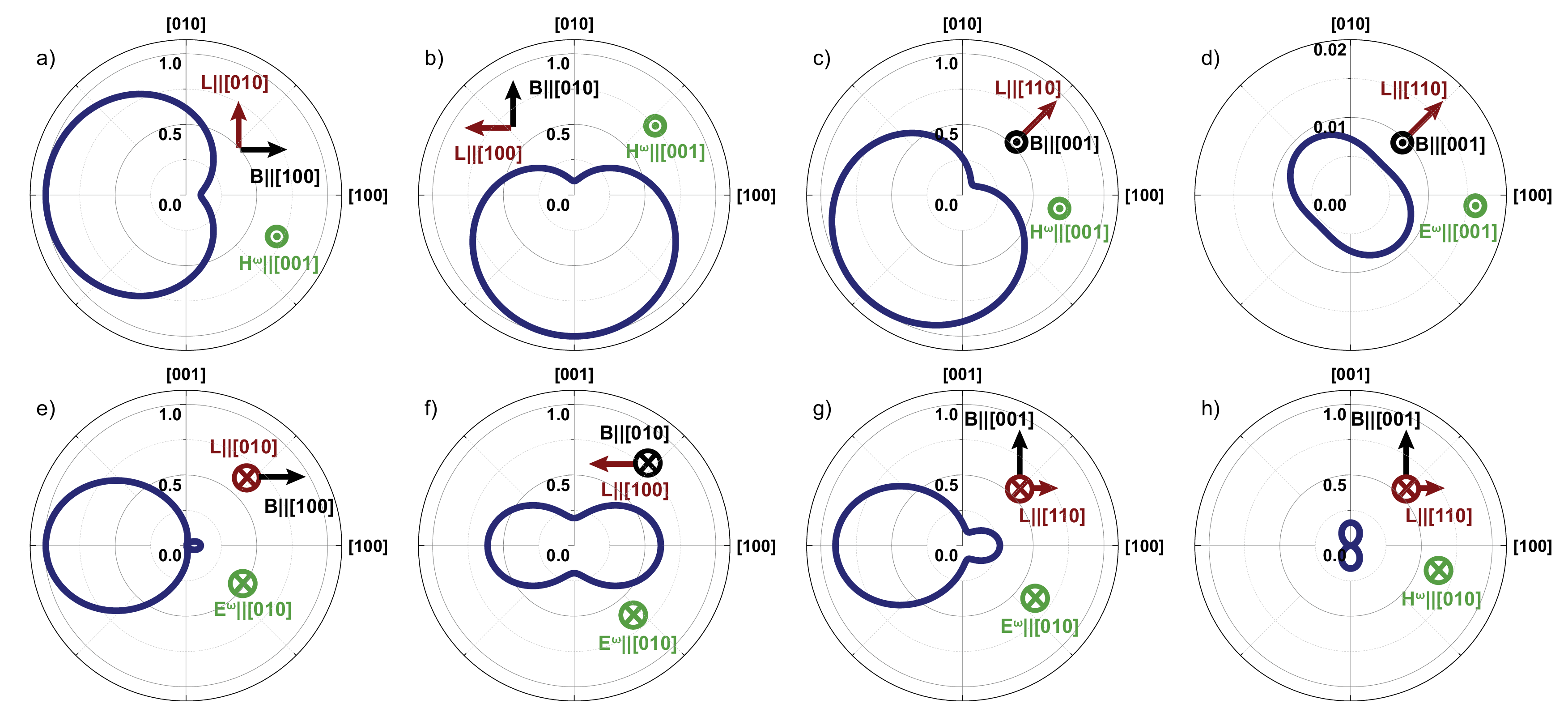}
  \caption{
Calculated angular dependence of the emission intensities from the components of the first excited doublet to the ground state under a fixed external magnetic field of $B=0.5$~T at $T=15$~K. For each diagram, the orientations of the magnetic field $\mathbf{B}$, the antiferromagnetic vector $\mathbf{L}$, and the fixed component of the emitted light wave are specified.
    The polar angle represents the orientation of the light wave vector $\mathbf{k}$ relative to the \textit{a} [100] crystal axis.
    The radial values reflect the normalized intensities calculated using Eq.~\eqref{eq:Ieg}.
   }
\label{fig:kRotations}     
 \end{figure*}

It may be more illustrative to plot the emission intensity distribution across wave vector angles for a fixed magnetic field orientation. Similar results are expected, as reversing the light wave vector is fundamentally equivalent to reversing the external magnetic field. The calculated diagrams for all principal geometries are shown in Fig.~\ref{fig:kRotations}. The geometries with $\mathbf{E^\omega} \parallel c$ are amended, since the $E_z$ component does not contribute to the electric-dipole transitions between the ground and the first excited states~\cite{Boldyrev2023}.

\begin{figure*}[hbt]
  \includegraphics[width=\linewidth]{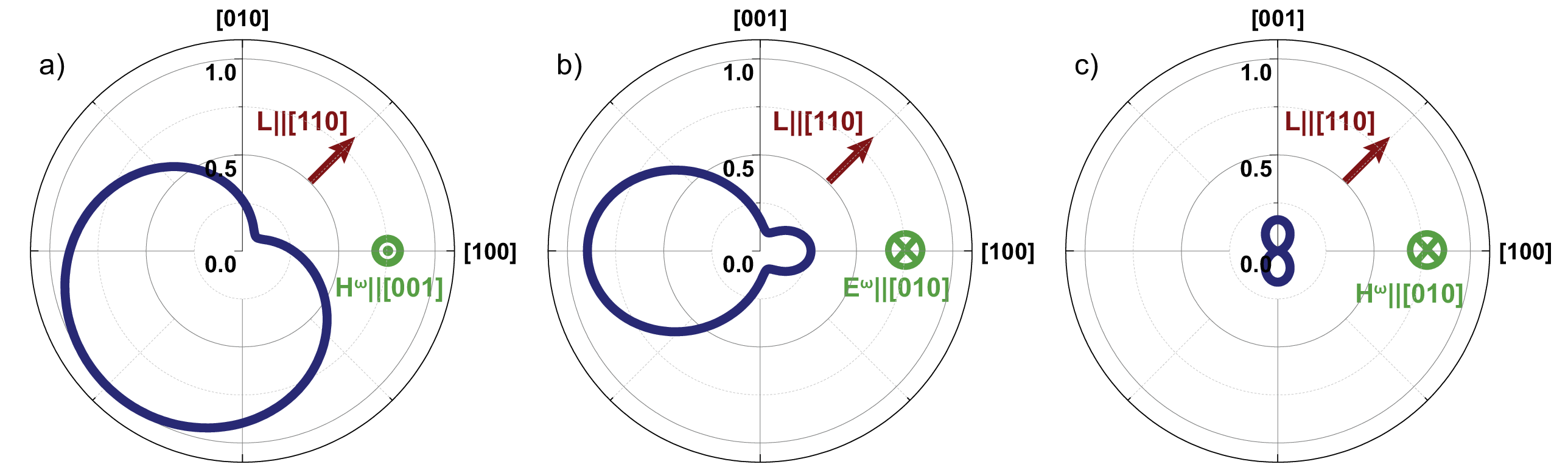}
  \caption{
    Calculated angular dependence of the emission intensities from the components of the first excited doublet to the ground state in the absence of an external magnetic field at $T=15$~K. The sample is assumed to be monodomain with the antiferromagnetic vector $\mathbf{L}\parallel[110]$ \cite{Boehm2003}. For each diagram, the orientations of the emitted light wave are specified. The polar angle represents the orientation of the light wave vector $\mathbf{k}$  relative to the \textit{a} [100] crystal axis.  The radial values reflect the normalized intensities calculated using Eq.~\eqref{eq:Ieg}.  
      }
\label{fig:kRotationsZeroField}     
 \end{figure*}

Notably, the nonreciprocity of emission can arise theoretically in a monodomain sample even in the absence of an external magnetic field, as it can be provided by the antiferromagnetic ordering of the Cu$^{2+}$ spins (Fig.~\ref{fig:kRotationsZeroField}). All diagrams in Figs. \ref{fig:kRotations} and \ref{fig:kRotationsZeroField} share the same normalized units, confirming that the effect intensity is independent of an external magnetic field. 

These results, in respect to symmetry selection rules, correspond to the phenomenological findings of Ref.~\onlinecite{Nikitchenko_2021} for the optical nonreciprocity, predicted for light transmission. One can see this by comparing Figs.~\ref{fig:kRotations} and \ref{fig:kRotationsZeroField} of the present study with the $\mathbf{kL}$ corrections in Figs.~9 and 5 from Ref.~\onlinecite{Nikitchenko_2021}.

The advantage of the microscopic approach that we developed here is in the deeper understanding of the intrinsic crystal properties. Within a single theoretical framework, it enables a unified analysis of polarization, absorption, and emission spectra. Furthermore, this approach inherently eliminates effects that might otherwise be considered phenomenologically feasible. A key illustration of this is the absence of nonreciprocity for the configuration $\mathbf{E^\omega} \parallel c$ in our case, despite the fact that the electric-dipole transitions are symmetry allowed.

Let us summarize the results of the theoretical consideration of the nonreciprocity of emission effect in a CuB$_2$O$_4$ crystal. Based on our microscopic theory, the nonreciprocity of emission in CuB$_2$O$_4$ originates from the interference between electric-dipole and magnetic-dipole transitions, which are of comparable magnitude. The primary role of the external magnetic field  is to control the orientation of the antiferromagnetic vector $\mathbf{L}$, thereby tuning the exchange interaction and modulating the interference conditions. This results in a strong directional dependence of the emission intensity on the relative orientations of the light wave vector, crystal axes, and magnetic field. The calculated emission diagrams predict geometries in which high emission nonreciprocity occurs, such as $\mathbf{k} {\parallel} a, \mathbf{B} {\parallel} a$ and $\mathbf{k} {\parallel} b, \mathbf{B} {\parallel} b$, whereas nonreciprocity is absent for $\mathbf{k} {\parallel} c$.

\clearpage

\section{Samples and Experimental Details}
\label{sec:Experimentals}

Single crystals of CuB$_2$O$_4$ were grown by the Kyropoulos technique from a melt of B$_2$O$_3$, CuO, Li$_2$O, and MoO$_3$ oxides~\cite{Petrakovskii2000}. To ensure a well-defined orientation of the samples, plane-parallel polished plates were cut from single crystals oriented using Laue X-ray diffraction.

\begin{figure}[hbt]
\begin{center}
\includegraphics[width=\linewidth]{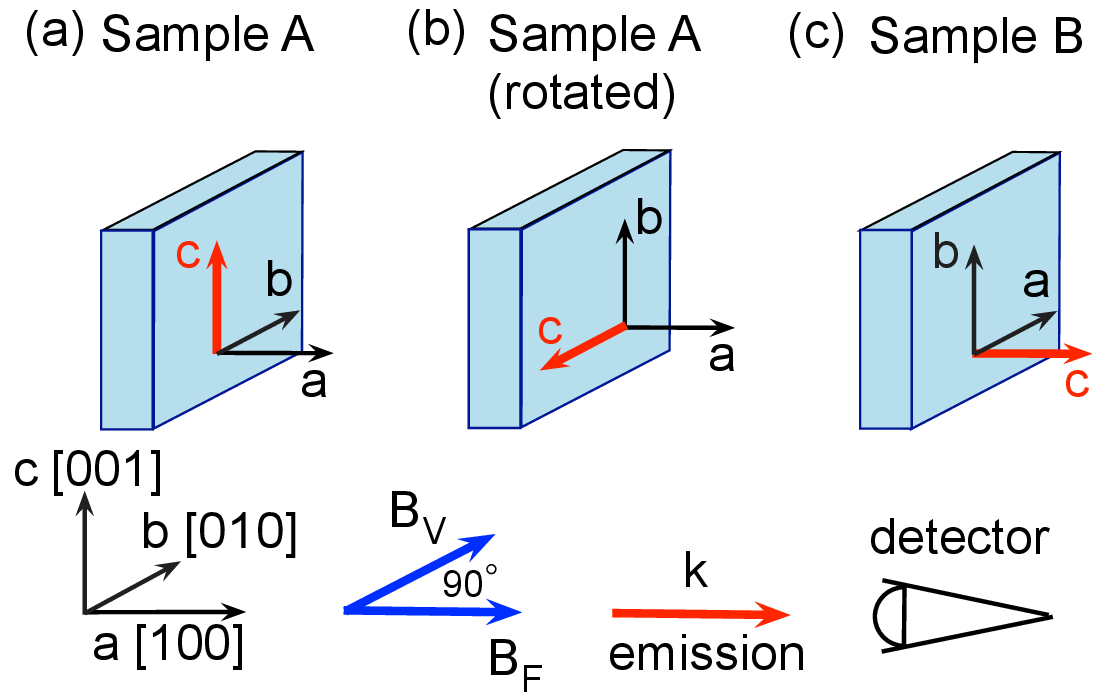}
\caption{Experimental geometries and orientation of the CuB$_2$O$_4$ samples A and B. The PL wavevector ($k$-vector) and the magnetic field directions in Faraday ($\textbf{B}_{\rm F} \parallel \textbf{k}$) and Voigt ($\textbf{B}_{\rm V} \perp \textbf{k}$) geometries are given at the bottom, they are the same for all crystal orientations. The upper panels show three orientations of the samples: (a) sample A with vertical $c$-axis, (b) rotated sample A with horizontal $c$-axis, and (c) sample B with out-of-plane $c$-axis. For the experiments with sample A, the emission linearly polarized perpendicular to the $c$-axis is measured ($\textbf{E} \perp c$).  
\label{geometries}
} 
\end{center}
\end{figure}

Two samples are studied. The sample A has a thickness of $1.12$~mm, the optical axis $c \parallel [001]$ is in the sample plane (Fig.~\ref{geometries}(a)). The $ab$ basal plane is perpendicular to the $[001]$ crystal direction and is spanned by the two $a$ and $b$ crystallographic axes ($a \parallel [100]$ and $b \parallel [010]$). Sample A was rotated in the $bc$ plane by 90$^\circ$ in order to obtain the geometries with either vertical (Fig.~\ref{geometries}(a)) or horizontal (Fig.~\ref{geometries}(b)) orientation of the $c$-axis. The sample B has a thickness of $429~\mu$m and its optical $c$-axis is oriented along the sample normal, $c \parallel [001]$ (Fig.~\ref{geometries}(c)).

The samples were mounted strain-free in a temperature-variable cryostat and the measurements were performed in the temperature range of $1.7-23$~K. For the lowest temperature of 1.7~K, the samples were immersed in pumped liquid helium. For temperatures exceeding 4.2~K, they were held in cold helium gas. For most of the reported experiments we used superconducting split-coil magnets with fields up to 10~T. The magnetic field was oriented either parallel to the $k$-vector of the emitted light (Faraday geometry, $\textbf{B}_{\rm F} \parallel \textbf{k}$) or perpendicular to the $k$-vector (Voigt geometry, $\textbf{B}_{\rm V} \perp \textbf{k}$). Some measurements were performed in a vector magnet with three superconducting coils oriented perpendicular to each other, so that any orientation of the magnetic field up to 3~T can be chosen.  

The photoluminescence (PL) was excited by a laser with photon energy of $E_\text{exc}=2.34$~eV, which strongly exceeds the detected emission of the lowest-in-energy $d-d$ transition of the $4b$ subsystem in the spectral range of $1.36-1.41$~eV. We chose this excitation to avoid possible effects related to changes of the laser absorption in magnetic field. Unpolarized laser light was used, while we checked that in case of linearly polarized light the emission intensity was insensitive to the orientation of the polarization plain, confirming that the light absorption is isotropic for strongly nonresonant excitation. Very low excitation densities of $P_\text{exc}=0.1$~mW/cm$^2$ were used in order to exclude possible heating effects, known for  CuB$_2$O$_4$~\cite{Kudlacik_2020}. Laser light was directed normal to the sample plane and PL was detected in backscattering geometry. Only for measuring signal in the $k_b B_b$ configuration we excited and collected the PL from the narrow side of the A-sample. The PL was detected in linear polarization, for the A-sample its stronger component perpendicular to the $c$-axis and for the B-sample, where emission is along the $c$-axis and unpolarized for vertical polarization. The emission was dispersed with an 0.5~m spectrometer and detected by a silicon charge-coupled-device camera with a spectral resolution of about 100~$\mu$eV. 

\section{Experimental results}
\label{Sec:Results}

The optical properties of CuB$_2$O$_4$ were examined by absorption~\cite{Pisarev2011}, photoluminescence~\cite{Toyoda2016,Kudlacik_2020}, Raman scattering~\cite{Pisarev2013}, and second harmonics generation~\cite{Pisarev2004}. Some of these results are discussed in an overview in Ref.~\cite{PisarevReview2023}. In our paper on low-temperature photoluminescence of CuB$_2$O$_4$ we suggested a classification of the emission lines contributing to the rich spectrum originating from Frenkel excitons~\cite{Kudlacik_2020}. 
In Ref.~\onlinecite{Toyoda2016}, the nonreciprocity of emission effect in CuB$_2$O$_4$ was first reported for only one experimental geometry and for magnetic fields up to 0.08~T. Our goal here is to present a comprehensive experimental picture of the  nonreciprocity of emission effect covering various geometries, temperatures, and magnetic fields. It is necessary to note, that such study is complicated by the rich magnetic phase diagram of CuB$_2$O$_4$ below the N\'eel temperature $T_{N}=20$~K, because it comprises a few commensurate and incommensurate phases. The phase diagram of the tetragonal CuB$_2$O$_4$ crystal is strongly anisotropic, and the phase transitions induced by an external magnetic field depend on the field strength and field orientation with respect to the crystal axes~\cite{Pankrats2018,Kudlacik_2024}.  

\subsection{Photoluminescence of CuB$_2$O$_4$ at different temperatures}
\label{PL}

\begin{figure}[hbt]
  \includegraphics[width=\linewidth]{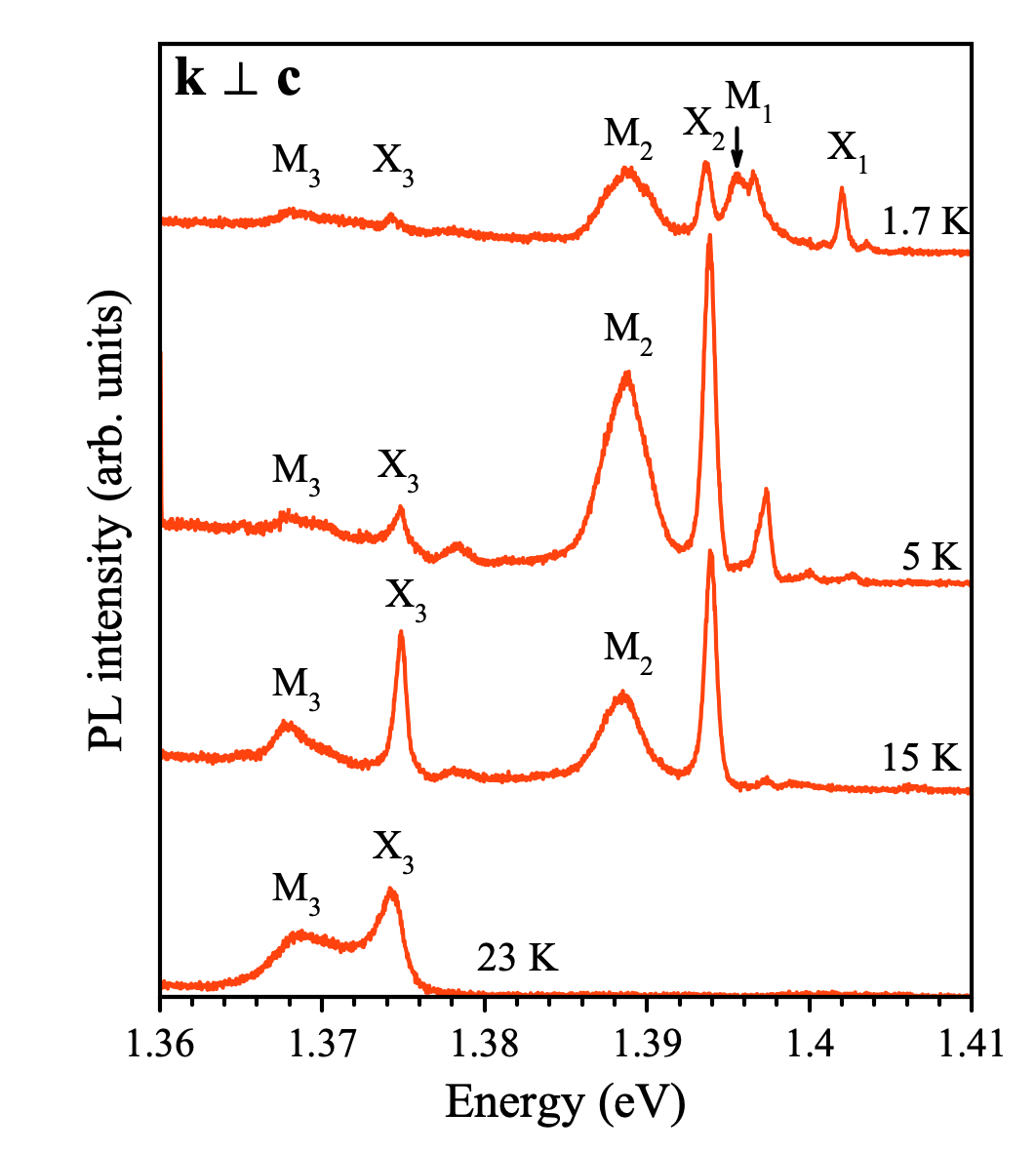}
  \caption{ Photoluminescence spectra of CuB$_2$O$_4$ (sample A) measured  at various temperatures for nonresonant excitation at $E_\text{exc}=2.34$~eV using $P_\text{exc}=0.1$~mW/cm$^2$. The exciton (X$_i$) and magnon (M$_i$) transitions in luminescence belonging to three different sets are labeled accordingly.
   }
\label{fig:PL}     
 \end{figure}

The low-temperature emission spectra of CuB$_2$O$_4$ are contributed by the Cu$^{2+}$ ions belonging to the $4b$ subsystem. In absorption, this subsystem is characterized by a narrow line at 1.4061~eV (at $T=1.6$~K) with a width of 0.6~meV and a peak, which was identified as a Frenkel exciton~\cite{Kudlacik_2020,Kopteva_2022}. PL spectra are shifted to lower energies from the absorption line. In Figure~\ref{fig:PL} we show PL spectra for the range of $1.36-1.41$~eV. They are complex showing many lines differing in width, intensity, recombination dynamics and temperature dependence. Their detailed study and suggested systematization of the contributing lines can be found in Ref.~\onlinecite{Kudlacik_2020}. 

In the PL spectrum measured at $T=1.7$~K one can see three sets of lines X$_i$ or M$_i$ ($i=1,2,3$). Here the X$_i$ are the exciton lines having different Stokes shifts relative to absorption: X$_1$ (1.4020~eV),  X$_2$ (1.3938~eV), and X$_3$ (1.3743~eV). The origin of the shifts has not yet been clarified. The Stokes shifts of 4.1~meV for X$_1$, 12.3~meV for X$_2$, and 31.8~meV for X$_3$ significantly exceed the linewidths, which allows us to exclude reabsorption as relevant factor for the nonreciprocity of emission measured in the present study. Note that the nonreciprocity of light propagation in exciton absorption was demonstrated in CuB$_2$O$_4$ experimentally~\cite{Toyoda2015,Toyoda2019} and analyzed theoretically in Ref.~\onlinecite{Nikitchenko_2021}. Each exciton line is accompanied by a magnon line M$_i$, which originates from exciton recombination assisted by magnon generation. The M$_i$ line is shifted from the exciton line to lower energy by about 6~meV and has a linewidth of about 4~meV provided by the magnon energy dispersion in the range of $2.6-7.9$~meV in CuB$_2$O$_4$~\cite{Martynov2004,Martynov2006,Kudlacik_2020}. 

Figure~\ref{fig:PL} shows that the intensities of the X$_i$M$_i$ sets change strongly with increasing temperature. At $T=1.7$~K the lines X$_1$M$_1$ and X$_2$M$_2$ are strong and have comparable intensities, while the X$_3$M$_3$ lines are weak. A temperature increase to only 5~K leads to disappearance of the X$_1$M$_1$ lines and the X$_2$M$_2$ lines become the strongest. At $T=15$~K the X$_3$M$_3$ lines gain intensity and become equal with the X$_2$M$_2$ lines. We have found that the nonreciprocity of emission is observed for all sets of X$_i$M$_i$ lines below $T_N$. However, above $T_N$ only the X$_3$M$_3$ lines are observed in the PL spectrum. In the following, we show experimental data for the strongest line set at the used temperature.  

\subsection{Nonreciprocity of emission at 15~K temperature}
\label{EN_15K}

To demonstrate the nonreciprocity of emission in various experimental geometries, we choose the X$_3$M$_3$ lines at $T~=~ 15$~K at which CuB$_2$O$_4$ is in the commensurate antiferromagnetic phase. In fact, several such phases exist with phase transitions between them induced by a magnetic fields of 0.1~T and 0.6~T for $\textbf{B}_c \parallel c$ and 20~mT and 50~mT for $\textbf{B}_a \perp c$~\cite{Pankrats2018,Kudlacik_2024}. 

\begin{figure*}[hbt]
  \includegraphics[width=\linewidth]{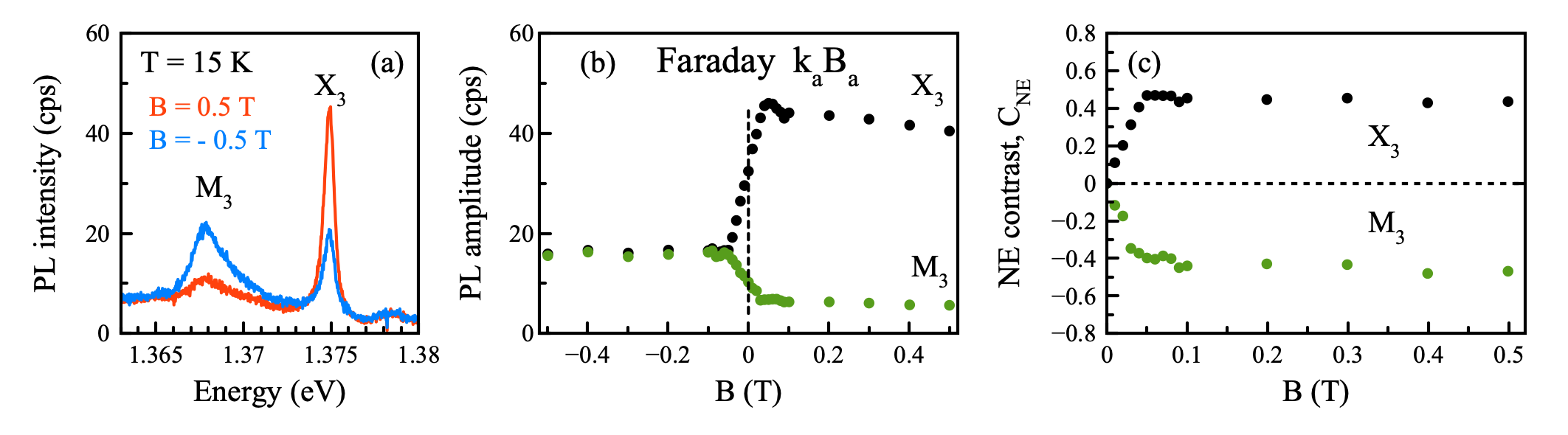}
  \caption{
Nonreciprocity of emission measured for the sample A at $T=15$~K for the $k_a B_a$ geometry ($\textbf{k} \parallel a$ and $\textbf{B}_{\rm F} \parallel a$) shown in Fig.~\ref{geometries}(a).
  (a) PL spectra at $B=0.5$~T (red) and $-0.5$~T (blue). 
  (b) PL amplitude of the X$_3$ and M$_3$ lines as function of magnetic field.  
  (c) Magnetic field dependence of the nonreciprocity contrast  for the X$_3$ and M$_3$ lines. 
      }
\label{fig:1Faa15K}     
 \end{figure*}

\begin{figure*}[hbt]
  \includegraphics[width=\linewidth]{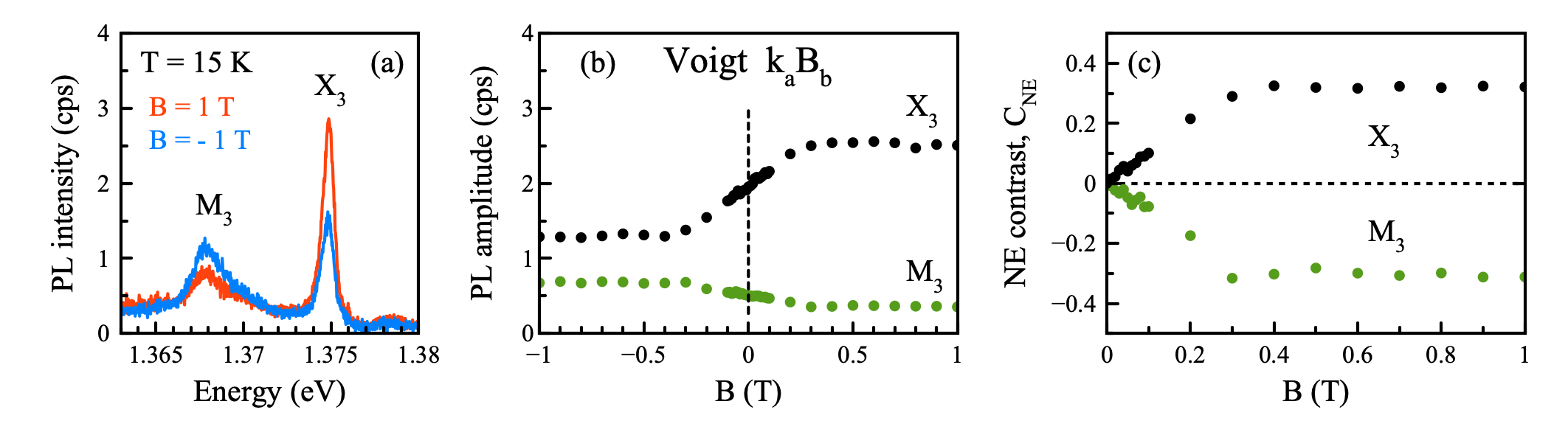}
  \caption{
  Nonreciprocity of emission for sample A measured at $T=15$~K for the $k_a B_b$ geometry ($\textbf{k} \parallel a$ and $\textbf{B}_{\rm V} \parallel b$) shown in Fig.~\ref{geometries}(a).
  (a) PL spectra at $B=1$~T (red) and $-1$~T (blue).
  (b) PL amplitude of the X$_3$ and M$_3$ lines as function of magnetic field.  
  (c) Magnetic field dependence of the nonreciprocity contrast  for the X$_3$ and M$_3$ lines. 
}
\label{fig:2Vab15K}     
 \end{figure*}

\begin{figure*}[hbt]
  \includegraphics[width=\linewidth]{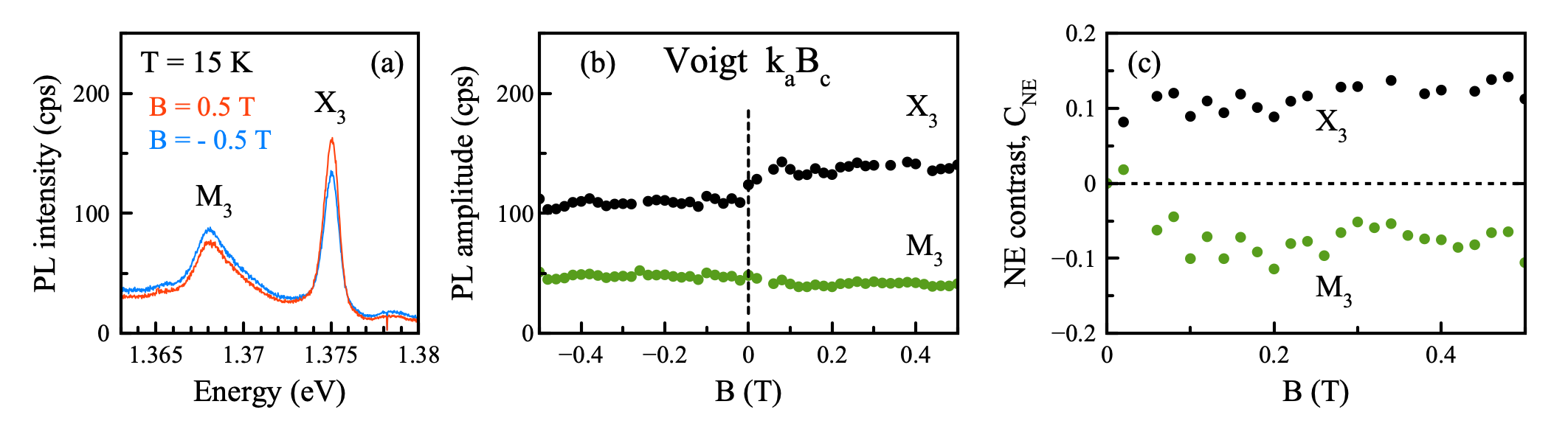}
  \caption{
  Nonreciprocity of emission measured for the rotated sample A at $T=15$~K for the $k_a B_c$ geometry ($\textbf{k} \parallel a$ and $\textbf{B}_{\rm V} \parallel c$) shown in Fig.~\ref{geometries}(b).
  (a) PL spectra at $B=0.5$~T (red) and $-0.5$~T (blue).
  (b) PL amplitude of the X$_3$ and M$_3$ lines as function of magnetic field.  
  (c) Magnetic field dependence of the nonreciprocity contrast  for the X$_3$ and M$_3$ lines. 
   }
\label{fig:3Vac15K}     
 \end{figure*}

\begin{figure*}[hbt]
  \includegraphics[width=\linewidth]{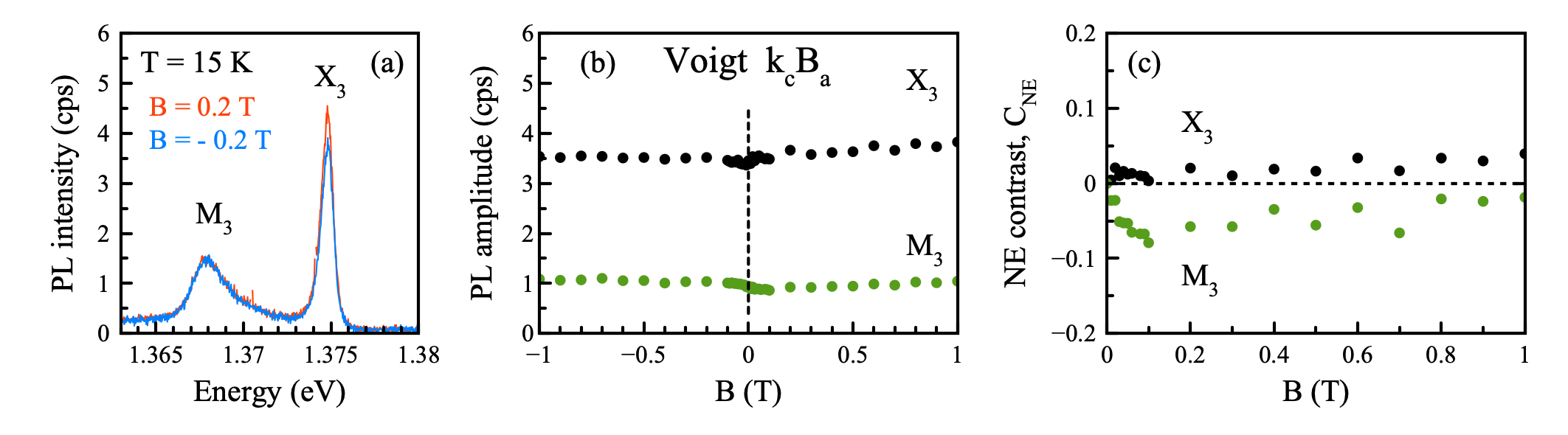}
  \caption{
  Nonreciprocity of emission measured for the sample B at $T=15$~K for the $k_c B_a$ geometry ($\textbf{k} \parallel c$ and $\textbf{B}_{\rm V} \parallel a$) shown in Fig.~\ref{geometries}(c).
  (a) PL spectra at $B=0.2$~T (red) and $-0.2$~T (blue).  
  (b) PL amplitude of the X$_3$ and M$_3$ lines as function of magnetic field.  
  (c) Magnetic field dependence of the nonreciprocity contrast  for the X$_3$ and M$_3$ lines. 
    }
\label{fig:7Vca15K}     
 \end{figure*}

\begin{figure*}[hbt]
  \includegraphics[width=\linewidth]{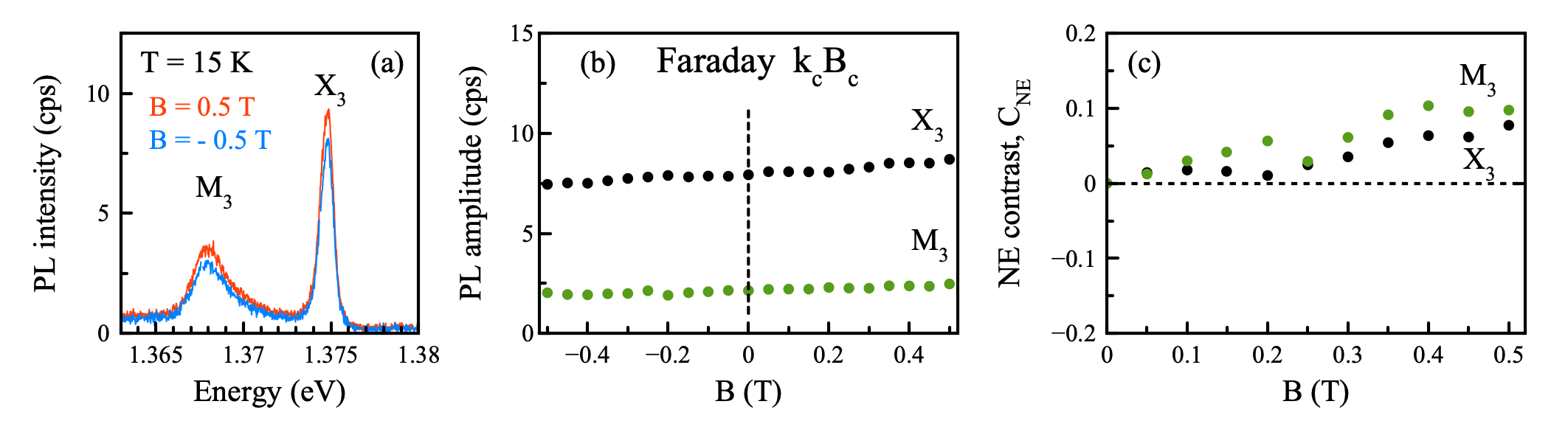}
  \caption{
  Nonreciprocity of emission measured for sample B at $T=15$~K for the $k_c B_c$ geometry ($\textbf{k} \parallel c$ and $\textbf{B}_{\rm F} \parallel c$) shown in Fig.~\ref{geometries}(c).
  (a) PL spectra at $B=0.5$~T (red) and $-0.5$~T (blue).
  (b) PL amplitude of the X$_3$ and M$_3$ lines as function of magnetic field.  
  (c) Magnetic field dependence of the nonreciprocity contrast  for the X$_3$ and M$_3$ lines.   
  }
\label{fig:9Fcc15K}     
 \end{figure*}

We have discussed and shown in the theoretical Sec.~\ref{sec:Theory} that similar properties are expected along the $a$- and $b$-crystal axes, which are equivalent in the tetragonal crystal. Therefore, there are five nonequivalent configurations to examine experimentally, namely $k_a B_a$, $k_a B_b$, $k_a B_c$, $k_c B_a$, and $k_c B_c$. The results for each of these configurations are shown in Figs.~\ref{fig:1Faa15K}, \ref{fig:2Vab15K}, \ref{fig:3Vac15K}, \ref{fig:7Vca15K}, and \ref{fig:9Fcc15K}, respectively. Each figure has three panels. Namely, PL spectra measured in magnetic fields of opposite directions are shown in panels (a) and magnetic field dependences of the PL peak intensity for the X$_3$ and M$_3$ lines in panels (b). Panels (c) show the nonreciprocity of emission contrast calculated as
\begin{equation}
\label{eq:CEN}
C_{\rm NE}=\frac{I(+B)-I(-B)}{I(+B)+I(-B)},
\end{equation}
where $I(+B)$ and $I(-B)$ are the PL intensities measured in opposite magnetic field directions. Below we present the experimental data and comment on their features. Their comparison with the model predictions of Sec.~\ref{sec:Theory} is given in Sec.~\ref{Sec:Discussion}.

In Figure~\ref{fig:1Faa15K}(a) the nonreciprocity of emission is shown for the X$_3$M$_3$ lines in the $k_{a} B_a$ geometry ($\textbf{k} \parallel a$ and $\textbf{B} \parallel a$) where it is expected to be strong, compare with the theory diagram in Fig.~\ref{fig:fieldDDL}(a) and Figs.~\ref{fig:kRotations}(a,e) for the horizontal line along the [100] axis. PL spectra are shown for opposite directions of the magnetic field. The intensity of the X$_3$ line is about twice stronger for the positive field direction (red spectrum), while for the M$_3$ line the effect is converse. This opposite change between the exciton X$_i$ and magnon M$_i$ lines is common also for other geometries and temperatures. It is also in good agreement with the theoretical predictions, see Fig.~\ref{fig:ddlX1M1}.   

The magnetic field dependences of the PL intensity of the X$_3$ and M$_3$ lines are given in Fig.~\ref{fig:1Faa15K}(b). The magnetic field is tuned from $-0.5$~T to $+0.5$~T. The intensities reach a constant level at higher fields, but change in weaker fields in the range of $-50$ to $+50$~mT. For quantitative presentation of the nonreciprocity of emission and its comparison for different conditions, it is convenient to use the nonreciprocity contrast defined by Eq.~\eqref{eq:CEN}. The magnetic field dependence of this contrast is shown in Fig.~\ref{fig:1Faa15K}(c). The opposite effect for the X$_3$ and M$_3$ lines, which saturate at 0.5 (meaning 50\%) and $-0.5$, respectively, is clearly seen. The nonreciprocity effect is absent at zero magnetic field, increases about linearly with growing field, and changes to saturation at 50~mT. Note, that this field strength corresponds to that of a magnetic phase transition between two commensurate phases (C$_0$ to C$_1$, we use here the phase denominations from Ref.~\onlinecite{Kudlacik2024}).  

In Figure~\ref{fig:2Vab15K} the $k_a B_b$ geometry case ($\textbf{k} \parallel a$ and $\textbf{B} \parallel b$), where nonreciprocity of emission is not expected (see theory diagram in Fig.~\ref{fig:fieldDDL}(a) for the vertical line along the [010] axis and Fig.~\ref{fig:kRotations}(f) for the horizontal line along the [100] axis), is presented. We see here a nonreciprocity with a contrast of 0.3 for the X$_3$ line and $-0.3$ for the M$_3$ line (Fig.~\ref{fig:2Vab15K}(c)). It increases much steeper with growing field, if to compare with the $k_a B_a$ geometry (Fig.~\ref{fig:1Faa15K}(c)) and turns to saturation at 0.3~T. This value cannot be assigned to a magnetic phase transition as investigated in Refs.~\onlinecite{Pankrats2018,Kudlacik_2024}.    

The $k_a B_c$ geometry case ($\textbf{k} \parallel a$ and $\textbf{B} \parallel c$) is illustrated in Fig.~\ref{fig:3Vac15K}. The nonreciprocity is present here, but its contrast is not large being 0.12 for the X$_3$ line and $-0.08$ for the M$_3$ line (Fig.~\ref{fig:3Vac15K}(c)). It saturates in a magnetic field of about $50-100$~mT, which can be associated with the magnetic phase transition from the C$^{c*}_0$ to the C$^{c}_1$ commensurate phases. Note that the theory gives no definite prediction for the nonrecipricity value in this case, as the result depends on the choice of the $\mathbf{L}$ direction. The diagram in Fig.~\ref{fig:ddlX1M1} predicts no effect for the vertical line along the [001] axis, while the finite nonreciprocity is predicted by the diagram in Figs.~\ref{fig:kRotations}(c,g) for the horizontal line along the [100] axis.   

In the $k_c B_a$ geometry ($\textbf{k} \parallel c$ and $\textbf{B} \parallel a$) the nonreciprocity is very small, so that one can assume that it is absent, see Fig.~\ref{fig:7Vca15K}. This is in agreement with the absence of the nonrecipricity in theory, see  Fig.~\ref{fig:fieldDDL}(c) for the horizontal line along the [100] axis and Fig.~\ref{fig:kRotations}(e) for the vertical line along the [001] axis.

Finally, in Fig.~\ref{fig:9Fcc15K} the $k_c B_c$ geometry ($\textbf{k} \parallel c$ and $\textbf{B} \parallel c$) is shown, where the nonreciprocity is not expected. The corresponding theory prediction is given in Fig.~\ref{fig:kRotations}(g) for the vertical line along the [001] axis. In experiment, one can see a weak linear increase of the nonreciprocity contrast, which saturate above 0.4~T. The saturation contrast values are 0.06 for the X$_3$ line and $0.10$ for the M$_3$ line. Note that both are positive. We rather consider this result as confirmation of the absence of nonreciprocity of emission for ideal experimental geometry and optimal crystal orientation, while other mechanisms not accounted for by the theory cannot be fully excluded. 

\begin{figure*}[hbt]
  \includegraphics[width=\linewidth]{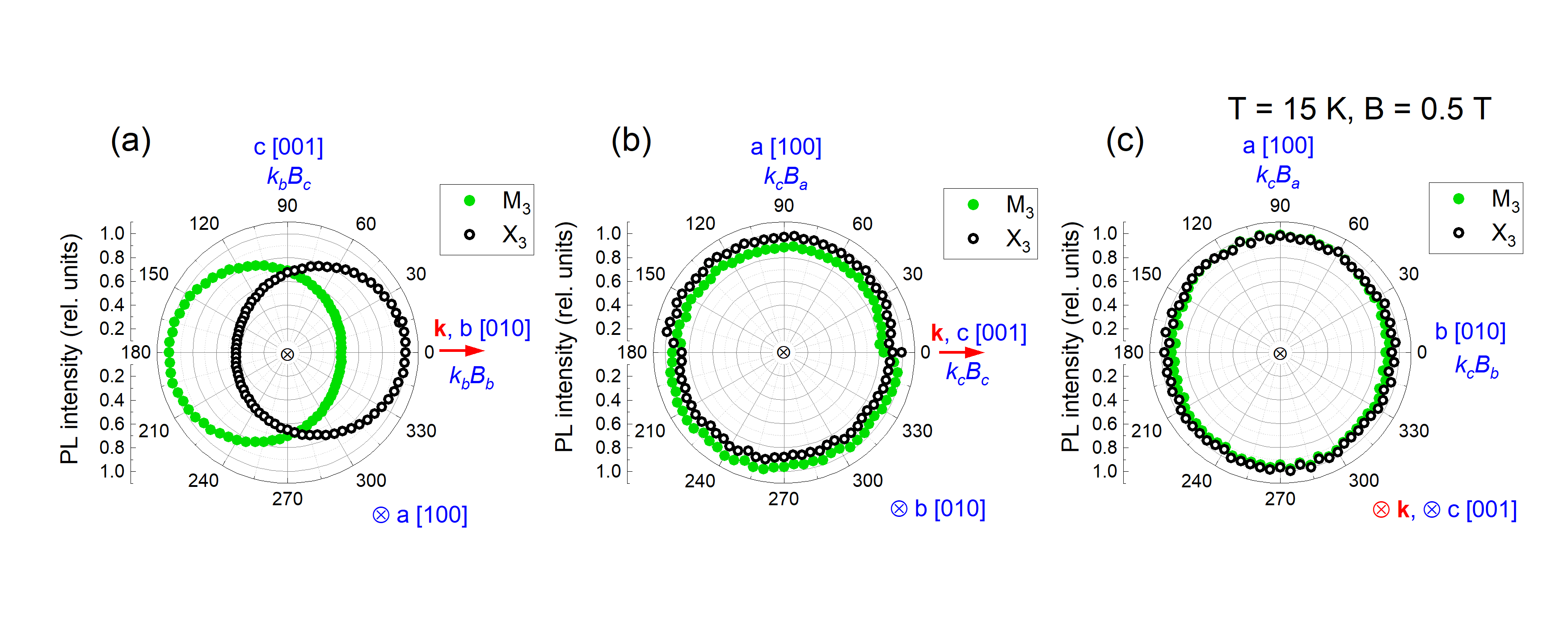}
  \caption{ Rotational diagrams for the PL intensity of the X$_3$ line (open black circles) and the M$_3$ line (closed green circles) measured at $T=15$~K in the vector magnet. The magnetic field of 0.5~T is rotated in various planes: (a) in the $(bc)$-plane for $\textbf{k} \parallel b$, (b) in the $(ac)$-plane for $\textbf{k} \parallel c$, and (c) in the $(ab)$-plane for $\textbf{k} \parallel c$. 
    }
\label{fig:Diagram_1_15K}     
 \end{figure*}

By using the vector magnet, we measure several rotational diagrams of the PL intensity for the X$_3$ and M$_3$ lines, see Fig.~\ref{fig:Diagram_1_15K}. The magnetic field is rotated in various planes: (a) in the $(bc)$-plane for $\textbf{k} \parallel b$, (b) in the $(ac)$-plane for $\textbf{k} \parallel c$, and (c) in the $(ab)$-plane for $\textbf{k} \parallel c$. These data are in agreement with our measurements in the Faraday and Voigt geometries for various sample orientations. They are also in agreement with the theory predictions in Fig.~\ref{fig:ddlX1M1} (for Fig.~\ref{fig:Diagram_1_15K}(a)) and in Fig.~\ref{fig:fieldDDL}(c) (for Figs.~\ref{fig:Diagram_1_15K}(b,c)).
 
\clearpage

\subsection{Nonreciprocity of emission at different temperatures}
\label{EN_dif_T}

It is instructive to study the nonreciprocity of emission at various temperatures and attribute its observation to specific magnetic phases. We remind that CuB$_2$O$_4$ at zero magnetic field is in the paramagnetic phase above the  N\'eel temperature of $T_N = 20$~K~\cite{Pankrats2018,Kudlacik_2024}. It is antiferromagnetically (AFM) ordered below $T_N$ and is in the commensurate AFM phase in the temperature range of $9-20$~K. Below 9~K it is in the incommensurate (IC) AFM phases (two of them are identified with the transition temperature of 5.8~K). Below 2~K there are two first-order phase transitions into a modulated state with wavevectors comparable with the lattice parameter~\cite{Pankrats2018,Lai2024}.

\begin{figure*}[hbt]
  \includegraphics[width=\linewidth]{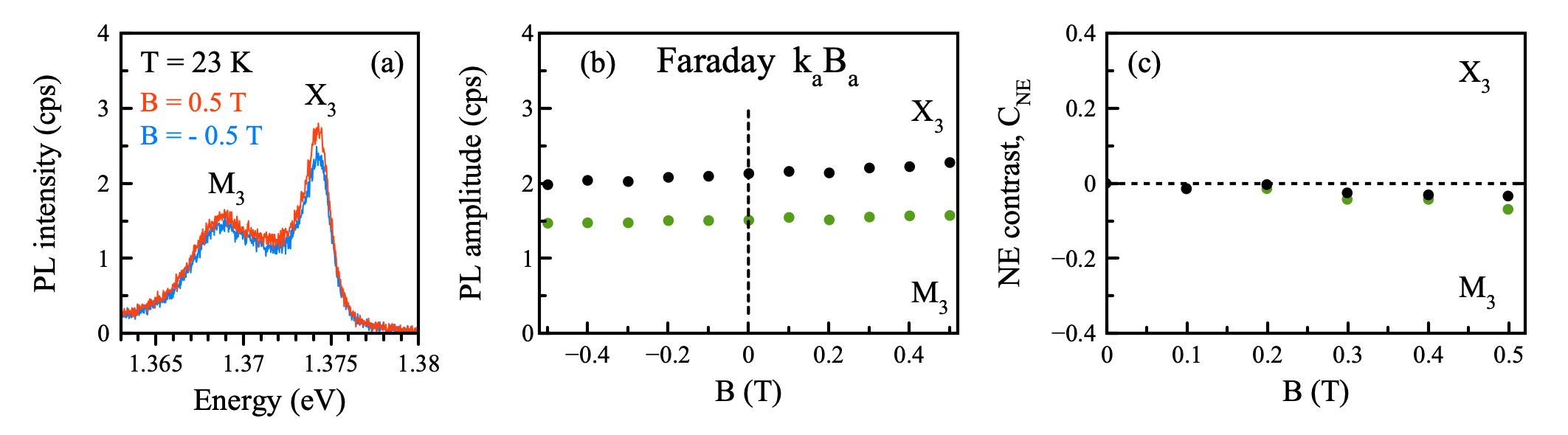}
  \caption{
  Nonreciprocity of emission measured for sample A at $T = 23$~K for the $k_a B_a$ geometry ($\textbf{k} \parallel a$ and $\textbf{B}_{\rm F} \parallel a$) shown in Fig.~\ref{geometries}(a).
  (a) Emission spectra at $B = 0.5$~T (red) and $-0.5$~T (blue). 
  (b) Emission amplitude of the X$_3$ and M$_3$ lines as function of magnetic field.  
  (c) Magnetic field dependence of the nonreciprocity contrast for the X$_3$ and M$_3$ lines. 
  }
\label{fig:1Faa23K}     
 \end{figure*}

For presenting the results at different temperatures, we chose the $k_{a} B_a$ geometry, where strong nonreciprocity was shown for $T=15$~K in Fig.~\ref{fig:1Faa15K}. We begin with the results at $T = 23$~K corresponding to the paramagnetic phase which are shown for the X$_3$M$_3$ lines in Fig.~\ref{fig:1Faa23K}. The nonreciprocity is absent for magnetic field varied in the range from $-0.5$~T to $0.5$~T. Therefore, we conclude that the nonreciprocity of emission is an inherent feature of the AFM ordering in CuB$_2$O$_4$. 

\begin{figure*}[hbt]
  \includegraphics[width=\linewidth]{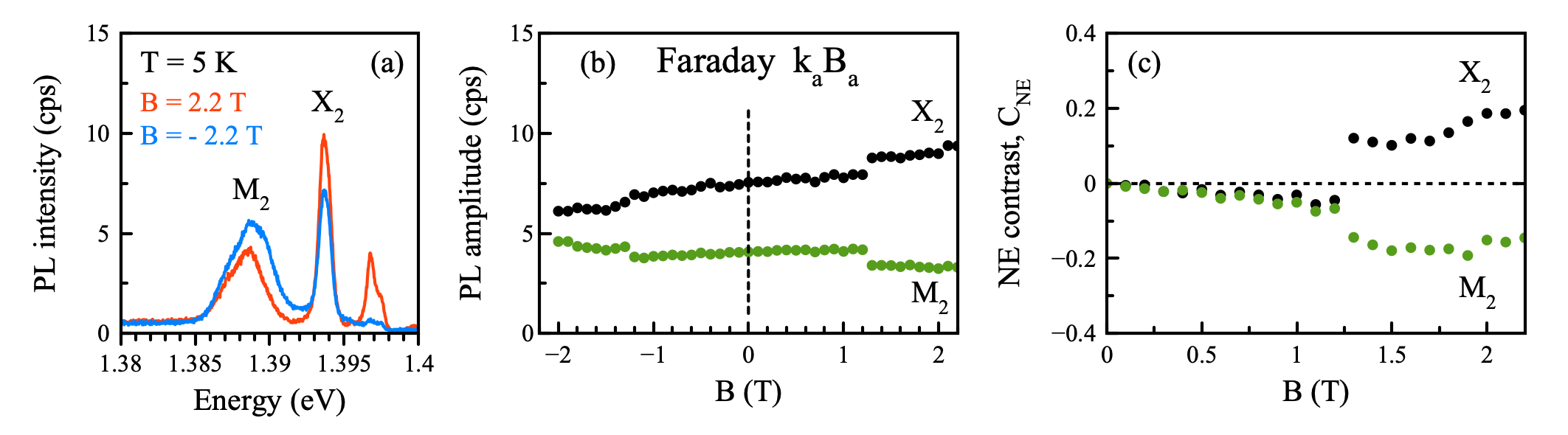}
  \caption{
  Nonreciprocity of emission measured for sample A at $T = 5$~K for the $k_a B_a$ geometry ($\textbf{k} \parallel a$ and $\textbf{B}_{\rm F} \parallel a$) shown in Fig.~\ref{geometries}(a).
  (a) Emission spectra at $B = 2.2$~T (red) and $-2.2$~T (blue). 
  (b) Amplitude of the X$_2$ and M$_2$ lines as function of magnetic field.  
  (c) Magnetic field dependence of the nonreciprocity contrast for the X$_2$ and M$_2$ lines. 
  }
\label{fig:1Faa5K}     
 \end{figure*}

The results for $T = 5$~K are shown in Fig.~\ref{fig:1Faa5K} for the X$_2$M$_2$ lines, which have large intensity at this temperature. Here, the nonreciprocity vanishes in magnetic fields below 1.25~T, namely when the sample is in the incommensurate AFM phase. It appears sharply in higher fields, when the phase transition from the IC$_2$ to the mixed C-IC phase occurs at 1.2~T~\cite{Kudlacik2024}. This result shows that in the incommensurate AFM phase the nonreciprocity of emission vanishes.

\begin{figure*}[hbt]
  \includegraphics[width=\linewidth]{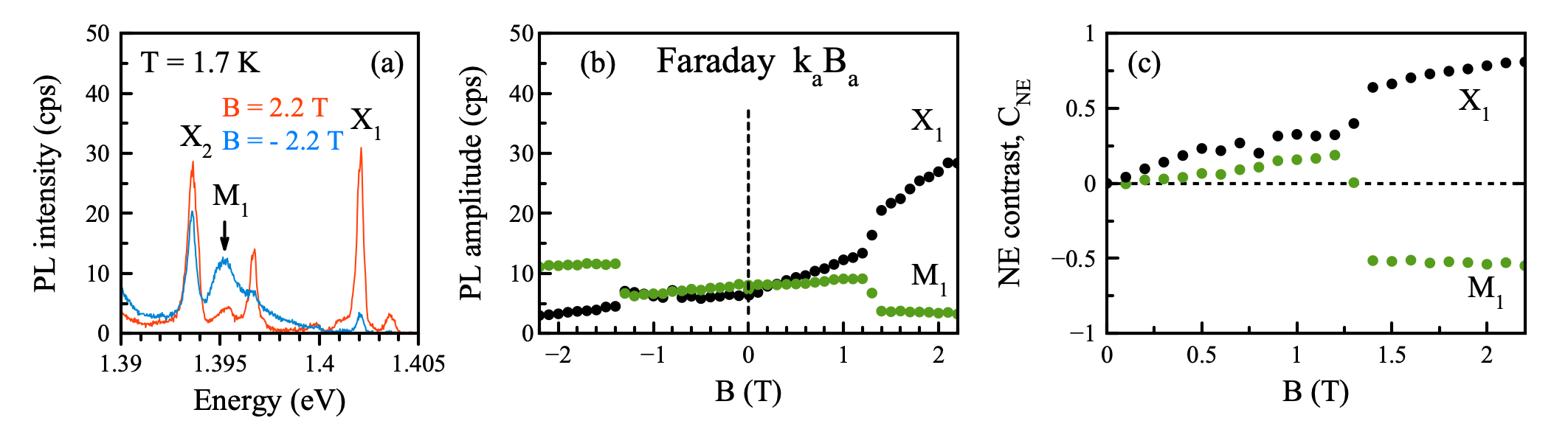}
  \caption{
  Nonreciprocity of emission measured for sample A at $T = 1.7$~K for the $k_a B_a$ geometry ($\textbf{k} \parallel a$ and $\textbf{B}_{\rm F} \parallel a$) shown in Fig.~\ref{geometries}(a).
  (a) Emission spectra at $B = 2.2$~T (red) and $-2.2$~T (blue). 
  (b) Emission amplitude of the X$_1$ and M$_1$ lines as function of magnetic field.  
  (c) Magnetic field dependence of the nonreciprocity contrast for the X$_1$ and M$_1$ lines. 
    }
  \label{fig:1Faa1.7K}     
 \end{figure*}

In Figure~\ref{fig:1Faa1.7K} the results for the X$_1$M$_1$ lines measured at $T = 1.7$~K are given. For the X$_1$ line, the nonreciprocity contrast increases linearly with field and reaches the value of $0.25$ at 0.5~T. Then it remains almost constant for a further field increase up to 1.35~T, where it jumps to $0.7$ and then slowly grows reaching $0.8$ at 2.2~T. The M$_1$ line shows, in general, a similar behavior. Its contrast increases linearly and reaches 0.15 at 1.2~T, and then changes sharply to the level of $-0.5$. Note that the field of 1.35~T corresponds to the transition from the IC$_2$ to the mixed C-IC phase~\cite{Kudlacik2024}.  Therefore, in line with other results, a strong nonreciprocity effect with opposite signs for the X$_1$ and M$_1$ lines is observed for the C-IC phase. For the IC$_2$ phase the effect is also observed, while it is positive for both lines. 

To conclude, the effect of nonreciprocity is demonstrated for the all X$_i$M$_i$ pairs of lines in the emission spectra of CuB$_2$O$_4$.  In turn, it is worth to note that the observed nonreciprocity effect provides an additional tool for identification of exciton and exciton-magnon emission lines.

\clearpage

\section{Discussion}
\label{Sec:Discussion}

\begin{figure*}[hbt]
  \includegraphics[width=0.8\linewidth]{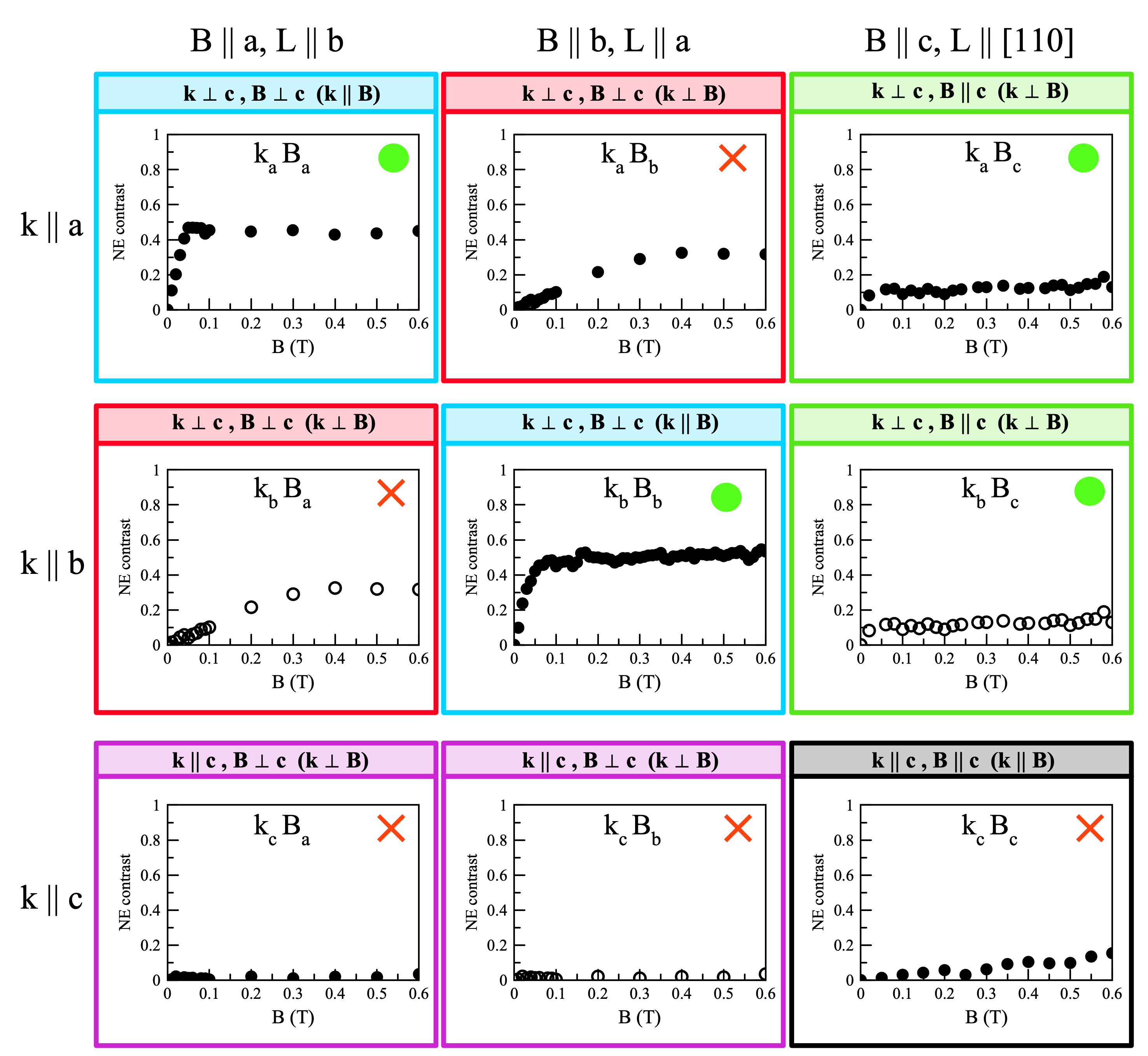}
  \caption{Overview of the experimentally measured nonreciprocity contrast for the X$_3$ line at $T=15$~K in different geometries. The data represented by the closed circles are measured data, while the open symbols represent data from the symmetry equivalent geometry. These geometries are marked by the same colors of the frame: $k_a B_a = k_b B_b$ (blue), $k_a B_b = k_b B_a$ (red), $k_a B_c = k_b B_c$ (green), and $k_c B_a = k_c B_b$ (violet). The green dots and the red crosses mark the geometries where the nonreciprocity of emission is allowed or forbidden theoretically, respectively. 
     }
\label{fig:Table2}     
 \end{figure*}

In order to summarize the experimental results measured in various geometries and compare them with the theoretical predictions, we plot in Fig.~\ref{fig:Table2} the nonreciprocity contrast measured at $T = 15$~K for the X$_3$ line. Here we show nine geometries with various orientations of the $c$-axis, the $k$-vector of emission and the external magnetic field. There are five equivalent pairs among them, which are highlighted by colors, because of the condition that the properties along the $a$- and $b$-axes are similar to each other. The measured experimental data are shown by the closed circles. By the open circles we show data, which are used to demonstrate the expected behavior for the equivalent geometries. The measured data are taken from Figs.~\ref{fig:1Faa15K}$-$\ref{fig:9Fcc15K}. For the $k_b B_b$ Faraday geometry, the data were obtained from the side of sample A with vertical orientation of the $c$-axis ($\textbf{k} \perp c$). As expected, these data are similar to those for the $k_a B_a$ geometry. 

The experimental data validate the main theoretical predictions. For easier comparison we label in Fig.~\ref{fig:Table2} the theoretically-allowed geometries with the green circles and the theoretically-forbidden geometries with the red crosses. One can see that the strong nonreciprocity reaching 0.5 (i.e. 50\%) is confirmed for the $k_a B_a$ and $k_b B_b$ Faraday geometries, where $\textbf{k} \perp c$. Note that the nonreciprocity contrast in this geometry reaches a value of 0.8 (i.e. 80\%) at $T = 1.7$~K, see Fig.~\ref{fig:1Faa1.7K}(c). The nonreciprocity vanishes for all geometries with the light propagating along the optical axis, $\textbf{k} \parallel c$ ($k_c B_a$, and, equivalently, $k_c B_b$), and $k_c B_c$. In the symmetry-allowed geometries $k_a B_c$ and $k_b B_c$ the effect is observed, but with a smaller value of about 0.12.

The only discrepancy between theory and experiment is found for the $k_a B_b = k_b B_a$ geometry, in which the effect is not expected. In the experiment, the emission contrast linearly increases with magnetic field and reaches a value of 0.3 for $B>0.4$~T.  We suggest that it is most probably caused by a non-ideal crystal orientation, which results in a leakage of the allowed polarizations. We have commented in Ref.~\onlinecite{Nurmukhametov2022} that nonreciprocity of emission could appear even for small mismatch angles of about $1-5^{\circ}$. Also our theoretical analysis of the microscopic mechanisms of the nonreciprocity of emission can be extended by accounting for the interference of the magnetic- and electric-quadrupole transitions including their modification by the external magnetic field. Such an analysis goes beyond the scope of the present study.

In CuB$_2${O}$_4$ the nonreciprocity of emission was found experimentally by Toyoda et al. in Ref.~\onlinecite{Toyoda2016}, where it was termed as direction-dependent luminescence effect. We are convinced that antiferromagnet nonreciprocity of light emission is the better choice of denomination for this effect, as it captures the physical mechanisms responsible for the nonreciprocity in  magnetically-ordered solids. In Ref.~\onlinecite{Toyoda2016} only one crystal orientation with face (-110) was examined with the motivation that the authors expected the most pronounced effect for it. The nonreciprocity contrast of 0.21 was reported there for the X$_3$ line at $T=15$~K. In our experiments at this temperature and for this line we measure the nonreciprocity contrast of 0.5 in the $k_a B_a$ geometry, which seems to be more favorable. In our studies we use much stronger magnetic fields than the 0.08~T in Ref.~\onlinecite{Toyoda2016}, which allows us to obtain a comprehensive picture of the nonreciprocity of emission effect in various magnetic phases. Toyoda et al. provided the correct suggestion on the origin of the nonreciprocity of emission in CuB$_2${O}$_4$ by attributing it to the interference of the electric-dipole and magnetic-dipole transitions. Our theoretical and experimental results presented here corroborate this suggestion.

\section{Conclusions}

We present experimental and theoretical studies of the nonreciprocity of light emission induced by antiferromagnetic ordering of Cu$^{2+}$ ions  in  CuB$_2${O}$_4$. Our comprehensive theoretical analysis and experimental results are made for various directions of the magnetic field, the wave vector of emission, and the polarization of the light wave. A very high nonreciprocity of emission reaching a contrast of 80\% for opposite field directions is observed for one of the experimental geometries. 
For all experimental geometries, a rigorous quantum mechanical calculations of the wave functions of the initial and final states of the Cu$^{2+}$ ion are made. Our microscopic calculations support the concept that the nonreciprocity of emission  in CuB$_2${O}$_4$ can be explained by the interference of magnetic-dipole and electric-dipole transitions within the magnetic Cu$^{2+}$ ions.

\textbf{Acknowledgments.} 
D.K., D.R.Y., and M.B. acknowledge financial support by the Deutsche Forschungsgemeinschaft through the Collaborative Research Center TRR142 (Project A11). M.V.E. acknowledges the subsidy allocated to Kazan Federal University for the state assignment in the sphere of scientific activities (Project No. FZSM-2024-0010). R.V.P. acknowledges the financial support from the RSF project 24-12-00348.

 \end{document}